\documentclass[12pt]{article}

\usepackage{latexsym,amsmath,amssymb,theorem,epsf}

\newcommand{\onefigure}[2]{\begin{figure}[htbp]
         \caption{\small #2\label{#1}(#1)}
         \end{figure}}
\renewcommand{\onefigure}[2]{\begin{figure}[htbp]
        \begin{center}\leavevmode\epsfbox{#1.eps}\end{center}
         \caption{\small #2\label{#1}}
      \end{figure}}

\DeclareFontFamily{OT1}{rsfs10}{} 
\DeclareFontShape{OT1}{rsfs10}{m}{n}{ <-> rsfs10 }{} 
\DeclareMathAlphabet{\mathscript}{OT1}{rsfs10}{m}{n} 

\numberwithin{equation}{section}

\newcommand{\cp}[1]{{\mathbb C}{\mathbb P}^{#1}}

\newcommand{\ns}{\normalsize}

\def\cB{{\mathcal B}}
\def\cC{{\mathcal C}}
\def\cE{{\mathcal E}}
\def\cF{{\mathcal F}}
\def\cG{{\mathcal G}}

\def\cL{{\mathcal L}}

\def\cM{{\mathcal M}}

\def\cO{{\mathcal O}}
\def\cP{{\mathcal P}}

\def\cS{{\mathcal S}}

\theoremstyle{plain} 

{\theorembodyfont{\rmfamily} }

\begin{document}


\begin{titlepage}

\vspace{-5cm}

\title{
   \hfill{\ns UPR-887T, OUTP-99-03P} \\[1em]
   {\LARGE Five--Brane BPS States in Heterotic M--Theory} \\[1em] } 
\author{
   Antonella Grassi$^1$, Zachary Guralnik$^2$ 
   and  Burt A.~Ovrut$^2$ \\[0.5em]
   {\ns $^1$Department of Mathematics, University of Pennsylvania} \\[-0.4em]
   {\ns Philadelphia, PA 19104--6395, USA}\\ 
   {\ns $^2$Department of Physics, University of Pennsylvania} \\[-0.4em]
   {\ns Philadelphia, PA 19104--6396, USA}\\}

\date{}

\maketitle

\begin{abstract}

We present explicit methods for computing the discriminant curves and the associated
Kodaira type fiber degeneracies of elliptically fibered Calabi--Yau
threefolds. These methods are applied to a specific three--family, $SU(5)$
grand unified theory of particle physics within the context of Heterotic
M--Theory. It is demonstrated that there is always a region of moduli space
where a bulk space five--brane is wrapped on a pure fiber in the Calabi--Yau
threefold. Restricting the discussion to the smooth parts of the discriminant
curve, we explore the properties of the $N=2$ BPS supermultiplets that 
arise on the worldvolume of this five--brane due to the degeneration of 
the elliptic fiber. The associated degenerating M membranes are shown
to project to string junctions in the base space. We use string junction techniques
to explicitly compute the light BPS hyper- and vector multiplet spectrum for each 
Kodaira type fiber near the smooth part of the discriminant curve in the
$SU(5)$ GUT theory.

\end{abstract}

\thispagestyle{empty}

\end{titlepage}


\section{Introduction:}


In a series of papers \cite{losw1, nse, don1,fbs, ms, ppm, si},
it was shown that when Ho\v rava--Witten theory \cite{HW1}
is compactified on Calabi--Yau threefolds, a ``brane Universe'' theory
\cite{losw1, dim} of particle physics, called Heterotic M--Theory, 
naturally emerges. This Universe consists of a
five--dimensional bulk space which is bounded by two four--dimensional BPS
three--branes, one at one end of the fifth dimension and one at the other. One
of these branes contains observed matter, such as quarks and leptons, as well
as their gauge interactions, either as a grand unified theory or as the
standard model. This is called the ``observable sector''. The other brane
can be chosen to contain pure gauge fields and their associated gauginos.
Gaugino condensates on this brane may provide a source of 
supersymmetry breaking in the theory. This brane is called the ``hidden
sector''.

As discussed in \cite{don1,ms, ppm}, 
the $E_{8}$ gauge group of Hor\v ava--Witten theory 
can be spontaneously broken to a GUT group or the
standard model gauge group on the observable brane by the appearance of a
semi--stable holomorphic vector bundle with structure group $G$ 
on the associated Calabi--Yau threefold. This bundle acts as a non--zero
vacuum expectation value, breaking the $E_{8}$ gauge group to the commutant 
$H$ of $G$. Commutant $H$ acts as the gauge group on
the observable brane. The physical requirements that $H$ be a realistic GUT
or standard model gauge group, that the theory be anomaly free and that there
be three families of quarks and leptons, puts strong constraints on the
allowed vector bundle. In general, there will be a solution if and only if one
further requires the appearance of five--branes, wrapped on holomorphic curves
in the associated Calabi--Yau threefold, located in the bulk space.

As shown in \cite{don1,ms,ppm}, the exact 
structure of the holomorphic curve on which these
bulk five--branes are wrapped can be computed from the anomaly cancellation
condition. In this paper, we will take the Calabi--Yau threefold, $X$, to
be an elliptic fibration over a base surface, $B$. 
In this case, it was shown in \cite{don1,ms,ppm}
that the holomorphic curve on which the bulk five--branes wrap generically
has both a base and a fiber component. However, as discussed in \cite{ms,si},
there are always regions of the associated moduli space where such a curve
decomposes into independent curves, at least one of which is a pure fiber.
We denote this pure fiber by $\cC_{2}$. When wrapped on this component, the
manifold of a five--brane is of the form $\cC_{2} \times M_{4}$. As is
well known, as long as the elliptic fiber is smooth, the $M_{4}$
worldvolume spectrum consists of an $N=4$ Abelian vector supermultiplet,
which is broken to an $N=1$ Abelian vector supermultiplet plus uncharged chiral
supermultiplets by higher dimensional operators in the effective theory.

However, not all fibers of an elliptic fibration are smooth. There are a
number of ways, classified by Kodaira \cite{kod}, in which the associated torii can
degenerate. The locus of all points in the base $B$ over which the fiber is
degenerate forms a divisor, called the discriminant curve. The structure
of this discriminant curve depends on the precise theory one is considering
and, in general, can be quite intricate. This curve generically has smooth
sections, cusps, and both normal and more complicated intersections. The
Kodaira type of degeneration of the fiber can change substantially from place
to place over the discriminant curve. If we choose some point on the
discriminant locus, and wrap the five--brane over the associated degenerate
elliptic fiber, then new physics emerges. If the point chosen is in a smooth
part of the curve, then the worldvolume theory on $M_{4}$ has $N=2$
supersymmetry at low energy. If one chooses the point at a singular locus of
the discriminant curve,  then the supersymmetry at low energies may be
further reduced to $N=1$.
In either case, new, massless states emerge, in addition to the
``standard'' worldvolume multiplets mentioned above. These new states arise
from the fluctuations of M membranes stretching between the vanishing cycles of
the torus. As the cycles go to zero, massless states emerge. 

The purpose of this paper is threefold. First, we present, for specificity, a
quasi--realistic, three--family $SU(5)$ grand unified theory within the context
of Heterotic M-Theory. This theory corresponds to an explicit semi--stable
holomorphic vector bundle with structure group $G=SU(5)$ over an elliptically
fibered Calabi--Yau threefold with base $B=\hat{\mathbb F}_{3}$. We discuss
the bulk five--branes and compute the class of the holomorphic curve over
which they are wrapped. The moduli space of this class is then presented and
it is shown that a region of this space corresponds to a single five--brane
wrapped on a pure elliptic fiber $\cC_{2}$. This work is presented in
Section $2$.
In Section $3$, we give the general theory for explicitly computing the
discriminant curves of elliptically fibered Calabi--Yau threefolds. We apply
these methods to determine the discriminant curves associated with the
specific
Calabi--Yau threefold with base $B=\hat{\mathbb F}_{3}$ presented in Section
$2$. We show that there are several possible curves, each with smooth
sections, cusps, and normal and tangential self--intersections. We compute the
Kodaira type fiber degeneracy over all points of the discriminant curve. In
the following section, Section $4$, we explicitly compute the $M_{4}$ worldvolume
BPS states that arise when a five--brane is wrapped on the degenerate fibers  
over the smooth parts of the discriminant curves. This reduces the problem
from one of two complex moduli to one modulus, and allows the application of 
standard Kodaira theory. Using the theory of string junctions developed in
\cite{bartonone,bartontwo}, we present the spectrum of BPS states, 
and the associated $N=2$ hyper- and vector multiplets,
for fibers of each Kodaira type occuring in the explicit theory of Section
$2$. The computation of light states 
over cusps and points of self--intersection, 
being inherently more intricate, 
will be presented in future publications \cite{next} . Finally, in the Appendix, 
we outline those aspects of string junction theory necessary for the
calculations in this paper. 

One can only speculate at this point about the possible physical role of the
BPS multiplets that arise in this manner. They will appear in the brane
world scenario as ``exotic'' charged matter living on the worldvolumes of
hidden sector three--branes. These new multiplets could be involved
in new mechanisms of supersymmetry breakdown \cite{fbs}, might be relevant to
cosmology \cite{cosm,cosm1,cosm2,cosm3,cosm4}, such as the dark matter in the
Universe and so on. We will return to these issues elsewhere.


\section{A Three Generation GUT Theory:}


In this section, we will construct a quasi--realistic particle physics
theory with three generations of quarks and leptons and a grand unified gauge
group SU(5). This is carried out within the framework of Heterotic M--Theory
compactified on an elliptically fibered Calabi--Yau threefold X. The
requirement that X be a Calabi--Yau manifold means that $c_{1}(TX)=0$, which,
in turn, implies that the base B of the elliptic fibration is restricted to be
a del Pezzo surface $d{\mathbb P}_{i}$, a Hirzebruch surface ${\mathbb
F}_{r}$, 
an Enriques
surface ${\bf E \rm}$ or certain ``blown--up'' ${\mathbb F}_{r}$ surfaces. 
For specificity, in 
this paper we will consider Calabi--Yau spaces with the base restricted to 
be of the latter type, that is, a blown--up Hirzebruch surface. As will 
become clear in the next section, we make this choice because elliptic 
fibrations over such a base have non--trivial discriminant curves involving 
not only different Kodaira type fibers, but also both normal and tangential 
crossing points. Again, for concreteness we will construct a model over a
specific blown--up Hirzebruch surface, although our results apply generally.

Consider the Hirzebruch surface ${\mathbb F}_{3}$. 
This is a ruled surface which is a
natural fibration of ${\mathbb F}_{3} \longrightarrow {\mathbb {CP}}^{1}$.  
We denote the fiber
class of this fibration by ${\cal{E}}$. This class has vanishing
self--intersection. In addition, there is a unique curve
of self--intersection $-3$ which we denote by  ${\cal{S}}$. These two classes
have a single point of intersection since ${\cal{E}} \cdot {\cal{S}} =1$. We
now modify this surface by blowing up a point on the curve ${\cal{E}}$ which
is not the point of intersection. The blow--up at that point introduces a new
exceptional class, which we denote as ${\cal{G}}$. This is the surface that we
will use as the base of our elliptically fibered Calabi--Yau threefold and we
denote it by
\begin{equation}
B=\hat{\mathbb F}_{3}
\label{eq:1}
\end{equation}
It is convenient to
introduce ${\cal{F}}$, where ${\cal{F}}+{\cal{G}}={\cal{E}}$. The three
classes
${\cal{S}}$, ${\cal{F}}$ and ${\cal{G}}$ are each effective classes and
together they form a basis of $H_{2}(\hat{\mathbb F}_{3}, {\mathbb Z} )$. 
Furthermore, they generate the Mori cone of effective classes. The
intersection numbers of these three classes are given by
\begin{equation}
{\cal{S}} \cdot {\cal{S}}= -3,  \qquad  {\cal{F}} \cdot {\cal{F}}=-1,
\qquad {\cal{G}} \cdot {\cal{G}}= -1
\label{eq:2}
\end{equation}
and
\begin{equation}
{\cal{S}} \cdot {\cal{F}}= 1,  \qquad  {\cal{S}} \cdot {\cal{G}}=0,
\qquad {\cal{G}} \cdot {\cal{F}}= 1
\label{eq:3}
\end{equation}
The first and second Chern classes of $\hat{\mathbb F}_{3}$ can be written as
\begin{equation}
c_{1}(\hat{\mathbb F}_{3})=2{\cal{S}}+5{\cal{F}}+4{\cal{G}}, \qquad 
c_{2}(\hat{\mathbb F}_{3})=5
\label{eq:4}
\end{equation}
respectively. Having specified the Chern classes of the base, we note that the
Chern classes of the tangent bundle, $TX$ of $X$ can now be computed. Since
$X$ is a Calabi--Yau threefold, $c_{1}(TX)=0$. However,  $c_{2}(TX)$ and
$c_{3}(TX)$ are found to be non-vanishing functions of 
$c_{1}(\hat{\mathbb F}_{3})$ and 
$c_{2}(\hat{\mathbb F}_{3})$. The exact expression for $c_{2}(TX)$ is given in
\cite{don1,FMW,cur,ba}.

We now want to specify a stable, holomorphic $SU(n)$ vector bundle over the
elliptically fibered Calabi--Yau threefold X with $B=\hat{\mathbb F}_{3}$. For
specificity, we will demand that the grand unification group be given by
\begin{equation}
H=SU(5)
\label{eq:5}
\end{equation}
which then requires that we choose the structure group of the vector bundle to
be
\begin{equation}
G=SU(5)
\label{eq:6}
\end{equation}
Hence, $n=5$. Having chosen $n$, the class of the spectral cover ${\cal{C}}$ 
of the vector
bundle is given by specifying a curve $\eta$ in the base $B=\hat{\mathbb F}_{3}$.
This curve must be effective to ensure that the spectral cover is
effective, as it must be. In addition, $\eta$ must be an irreducible curve
so that the associated vector bundle will be stable. Since
${\cal{S}}$, ${\cal{F}}$ and ${\cal{G}}$ are a basis of 
$H_{2}(\hat{\mathbb F}_{3}, {\mathbb Z} )$, we can always write
\begin{equation}
\eta=a{\cal{S}}+b{\cal{F}}+c{\cal{G}}
\label{eq:7}
\end{equation}
where $a$, $b$ and $c$ are integers. Recalling that the
classes ${\cal{S}}$, ${\cal{F}}$ and ${\cal{G}}$ generate the Mori 
cone, it follows that the condition for $\eta$ to be an effective  
class is simply that
\begin{equation}
a \geq 0, \qquad b \geq 0, \qquad c \geq 0
\label{eq:8}
\end{equation}
The conditions for the irreducibility of $\eta$ are a little more intricate to
derive. Here we simply state the result, which is that either
\begin{equation}
a \neq 0, \quad b =c=0  \qquad b \neq 0, \quad a =c=0  \qquad c \neq 0,
\quad a
=b=0
\label{eq:9}
\end{equation}
or
\begin{equation}
b \geq a, \qquad b \geq c, \qquad a \geq b-c
\label{eq:10}
\end{equation}
Having chosen the spectral cover ${\cal{C}}$ by giving $n$ and $\eta$, it
is now
necessary to specify the spectral line bundle ${\cal{N}}$ over ${\cal{C}}$.
Generically, the allowed line bundles are indexed by a rational number
$\lambda$. For $n$ odd, which is the case in our theory, 
this parameter must satisfy $\lambda \in {\mathbb Z}+\frac{1}{2}$. For
specificity, we will choose
\begin{equation}
\lambda=\frac{1}{2}
\label{eq:11}
\end{equation}
Having specified  ${\cal{C}}$ and ${\cal{N}}$ subject to the above conditions,
a stable, holomorphic $SU(5)$ vector bundle $V$ can be constructed using the
Fourier--Mukai transformation
\begin{equation}
({\cal{C}},{\cal{N}}) \longrightarrow V
\label{eq:12}
\end{equation}
We need not discuss $V$ here, other than to say that its properties can be
calculated from the above data. In particular, in addition to its vanishing
first Chern class, $c_{1}(V)=0$, the second and third Chern classes,
$c_{2}(V)$
and $c_{3}(V)$ respectively, can be computed and are found to be functions of
$n$, $\eta$ and $\lambda$. The exact expressions are given in
\cite{don1,FMW,cur,ba}.

As discussed in \cite{don1}, the physical requirement that the theory 
be anomaly free leads to the topological expression
\begin{equation}
W=c_2(TX)-c_{2}(V)
\label{eq:13}
\end{equation}
$W$ is a class that is interpreted as being a holomorphic curve in the
Calabi--Yau
threefold $X$ on which five--branes, located in the five--dimensional bulk
space, are wrapped. Using the exact expressions for $c_{2}(TX)$ and 
$c_{2}(V)$, we can compute this five--brane class explicitly. We
find that
\begin{equation}
W=W_{B}\sigma+a_{f}F
\label{eq:14}
\end{equation}
where $\sigma$ is the class of the zero section of the elliptic fibration $X$
and $F$ is the generic class of its fiber. Furthermore,
\begin{equation}
W_{B}= 12c_{1}(B)-\eta
\label{eq:15}
\end{equation}
and 
\begin{equation}
a_{f}=c_{2}(B)+(11+\frac{n^{3}-n}{24})c_{1}(B)^{2}-\frac{3n}{2\lambda}(\lambda^{2}
-\frac{1}{4})
\label{eq:16}
\end{equation}
For the above specific theory, both $W_{B}$ and $a_{f}$ can be computed and
are given by
\begin{equation}
W_{B}=(24-a){\cal{S}}+(60-b){\cal{F}}+(48-c){\cal{G}}
\label{eq:17}
\end{equation}
and
\begin{equation}
a_{f}=117
\label{eq:18}
\end{equation}
respectively, where we have used equations (\ref{eq:4}), (\ref{eq:7}) 
and (\ref{eq:11}), as well as $n=5$, $\lambda=\frac{1}{2}$ and 
equations (\ref{eq:2}), (\ref{eq:3}). As discussed 
in \cite{don1}, the class $W$ is
further constrained by the requirement that it be an effective class in
$H_{2}(X, {\mathbb Z})$. It was shown that this will be the case if and
only if $W_{B}$ is an effective class in $H_{2}(\hat{\mathbb F}_{3},
{\mathbb Z})$
and $a_{f}$ is a non--negative integer. It follows from (\ref{eq:18}) that the
last condition is satisfied. We see from equation (\ref{eq:17}) that $W_{B}$
and, therefore, $W$ will be an effective curve if and only if
\begin{equation}
24 \geq a, \qquad 60 \geq b, \qquad 48 \geq c
\label{eq:19}
\end{equation}
In addition to the anomaly cancellation condition (\ref{eq:13}), there is
another property required of any realistic theory of particle physics, that
is, that the number of quark and lepton generations be 3. As discussed in
\cite{ don1,cur},
the number of quark and lepton generations is given by
$N_{gen}=\frac{c_{3}(V)}{2}$. Using the expression for $c_{3}(V)$, it was
shown that the three family condition imposes the further constraint that
\begin{equation}
\lambda(W_{B}^{2}-(24-n)W_{B} \cdot c_{1}(B)+ 12(12-n)c_{1}(B)^{2})=3
\label{eq:20}
\end{equation}
For the above specific theory, using equations (\ref{eq:4}) and (\ref{eq:17}), as
well as $n=5$ and the intersections in (\ref{eq:2}), (\ref{eq:3}), this
condition becomes
\begin{equation}
-3a^{2}+2ab-b^{2}+2bc-c^{2}+5(a-b-c)=6
\label{eq:21}
\end{equation}
Therefore, to get a realistic $SU(5)$ grand unified theory with three families
of quarks and leptons for elliptically fibered Calabi--Yau threefolds with
base $B=\hat{\mathbb F}_{3}$, we must find a curve $\eta$ of form (\ref{eq:7})
that satisfies the conditions (\ref{eq:8}),(\ref{eq:9}) or (\ref{eq:10}),
(\ref{eq:19}) and (\ref{eq:21}) simultaneously.

The solution of these constraints is not entirely trivial. For the purposes of
this paper, we will only give the simplest solution. We find that it is
impossible to find any solutions for curves $\eta$ satisfying the conditions
in equation (\ref{eq:9}). We must, therefore, impose equation (\ref{eq:10}).
As an ansatz, we try for solutions with $a \neq 0$ and $b=c \neq 0$. Under
these conditions, we find a solution with 
\begin{equation}
a=6, \qquad b=c=42
\label{eq:22}
\end{equation}
corresponding to the curve
\begin{equation}
\eta= 6{\cal{S}}+42{\cal{F}}+42{\cal{G}}
\label{eq:23}
\end{equation}
and the five--brane class $W=W_{B}\sigma +117F$, where
\begin{equation}
W_{B}=18{\cal{S}}+18{\cal{F}}+6{\cal{G}}
\label{eq:24}
\end{equation}
The bulk space five--branes of this specific theory, described by the class
$W= W_{B}\sigma +117F$ with $W_{B}$ given in (\ref{eq:24}), are the objects of 
interest in this paper.

To analyze the physical structure of these five--branes, it is useful to first
construct their moduli space. The moduli spaces of M-theory five--branes
wrapped on holomorphic curves in elliptically fibered Calabi--Yau threefolds
were constructed in \cite{ms}. Here, we will simply apply the results of 
\cite{ms} to the 
specific theory discussed above. We find that the moduli space of these
five--branes is given by
\begin{equation}
{\cal{M}}(W_{B}\sigma+117F)= \bigcup^{117}_{n=0}{\cal{M}}_{0}
(W_{B}\sigma+nF)\times {\cal{M}}((117-n)F)
\label{eq:25}
\end{equation}
where $W_{B}$ is given in (\ref{eq:24}) and
\begin{equation}
{\cal{M}}((117-n)F)= (\hat{\mathbb F}_{3} \times {\mathbb
{CP}}^{1}_{a})^{117-n}/
{\mathbb Z}_{117-n}
\label{eq:26}
\end{equation}
${\mathbb {CP}}^{1}_{a}$ is the moduli space of the translation/axion
multiplet associated with the fifth direction of the bulk space.
The components ${\cal{M}}_{0}(W_{B}\sigma+nF)$ of the 
moduli space can also be explicitly computed. This construction was 
presented in \cite{ms}, where specific examples were given. It was shown that 
these components, which can be rather complicated, depend sensitively 
on the choice of the
base surface of the elliptic fibration and on the explicit form of the curve 
$W_{B}$  . However, in this paper, it is not necessary to know the explicit
form of these components of moduli space, and we will not discuss them
further.

We are particularly interested in the component of moduli space where all
the fibers are associated with the base curve $W_{B}$ except for one, that
is, when
\begin{equation}
n=116
\label{eq:27}
\end{equation}
The relevant component of moduli space is then
\begin{equation}
{\cal{M}}_{0}(W_{B}\sigma+(116)F)\times {\cal{M}}(F)
\label{eq:28}
\end{equation}
where
\begin{equation}
{\cal{M}}(F)=\hat{\mathbb F}_{3} \times {\mathbb {CP}}^{1}_{a}  
\label{eq:29}
\end{equation}
Physically, this region of moduli space describes a situation where all the
five--branes are wrapped on the complicated curve $W_{B}\sigma +116F$
except for one,
which is wrapped on a vertical curve described by 
the pure fiber class $F$. The moduli of this single
five--brane, specified by the space (\ref{eq:29}), are independent of the
all other
moduli. Therefore, this single five--brane can be discussed entirely by
itself, without reference to the rest of the five--brane class. In the
remainder of this paper, we will consider only this single five--brane
wrapped on
a curve described by the fiber class $F$. If, furthermore, 
we hold this five--brane fixed at a 
point $y_{0}$ in the fifth direction of the bulk space, the moduli space
is reduced to
\begin{equation}
{\cal{M}}(F)=\hat{\mathbb F}_{3}
\label{eq:30}
\end{equation}
The physical properties of this single five--brane are then completely
determined by the behavior of the elliptic fiber $F$ on which the five--brane
is wrapped as one moves around the base space $B=\hat{\mathbb F}_{3}$. 

The fiber over a generic point in $B=\hat{\mathbb F}_{3}$ is a smooth torus.
Therefore, the worldvolume fields of a five--brane wrapped on a generic fiber
consist of the standard ones associated with the self--dual anti--symmetric
tensor
multiplet. However, as is well known, the torus fibers can become singular at
specific points in the base surface. The locus of points where the fibers 
degenerate is a divisor of the base called the discriminant locus. That is, 
the discriminant locus is a smooth curve in $B=\hat{\mathbb F}_{3}$. Let us
choose a point on the discriminant curve and wrap a five--brane around the
degenerate torus fibered over that point. Then, as is well known, in addition
to the usual fields associated with the self--dual anti--symmetric tensor
multiplet, there are also massless BPS multiplets that appear on the five--brane
worldvolume. The properties of these new states are directly 
related to the ``singularity structure'' of the elliptic fiber at that point, 
which has been classified by Kodaira \cite{kod} and will be discussed below. 
The ensemble of these new states form a conformal field theory whose
construction
will be one of the goals of this paper.

Furthermore, the structure of the torus degeneration, that is, the Kodaira
type of the elliptic fiber, can change as one
considers different points on the discriminant curve. Therefore, the conformal
field theory that appears when a five--brane is wrapped on a degenerate torus
over one point of the discriminant curve, can be very different from that
which occurs when it is wrapped on a degenerate fiber over a
different point on the curve. In this paper, we will present the general
theory
for determining the exotic multiplets and the associated conformal field
theories on the five--brane worldvolume.
It is clear that the starting point of our analysis must be the
construction of the
allowed discriminant curves in the base $B=\hat{\mathbb F}_{3}$, and the 
computation of the exact Kodaira structure of the fiber degeneration. This
will be
carried out in the next section.


\section{Constructing The Discriminant Curves:}


A simple representation of an elliptic curve is given in the 
projective space $\cp{2}$ by the Weierstrass equation
\begin{equation}
zy^2=4x^3-g_2xz^2-g_3z^3
\label{eq:31}
\end{equation}
where $(x,y,z)$ are the homogeneous coordinates of $\cp{2}$ and $g_2$,
$g_3$ are constants. The origin of the elliptic curve is located at
$(x,y,z)=(0,1,0)$. The torus described by~\eqref{eq:31} can become degenerate
if one of its cycles shrinks to zero. Such singular behavior is characterized
by the vanishing of the discriminant
\begin{equation}
\Delta= g_{2}^{3}-27g_{3}^{2}
\label{eq:31A}
\end{equation}

Equation~\eqref{eq:31} can also represent an elliptically 
fibered threefold, $W$, if the coefficients $g_2$ and $g_3$ in 
the Weierstrass equation are functions over a base surface, $B$. 
Clearly $W$ constructed in this way 
has a zero section $\sigma$ and we denote by ${\cal{L}}$ the co--normal
bundle 
over $\sigma$. The correct way to express this fibration
globally is to replace the projective plane 
$\cp{2}$ by a fourfold $\cp{2}$-bundle $P \to B$ where $P = {\mathbb
P}({\mathcal
O}_{B}\oplus \cL^{2} \oplus \cL^{3})$. The  notation ${\mathbb P}(M)$ stands
for the projectivization of a vector bundle $M$. There is a hyperplane line
bundle ${\mathcal O}_{P}(1)$ on $P$ which corresponds to the divisor 
${\mathbb P}(\cL^{2}\oplus \cL^{3}) \subset P$ and the
coordinates $x,y,z$ are sections of $\cO_{P}(1)\otimes\cL^2,
\cO_{P}(1)\otimes
\cL^3$ and $\cO_{P}(1)$ respectively. Equation (\ref{eq:31}) can now be
interpreted as the affine form of a global equation on $P$ involving the
sections $x,y,z$, as long
as we require $g_{2}$ and $g_{3}$ to be sections of appropriate line 
bundles over the base $B$. It follows from \eqref{eq:31} 
that 
\begin{equation}
g_2 \sim \cL^4, \qquad 
g_3 \sim \cL^6
\label{eq:32}
\end{equation}
The symbol ``$\sim$'' simply means ``section of''. The zero locus of
equation (\ref{eq:31}) defines an elliptically fibered threefold hypersurface
of $P$, which is called a Weierstrass fibration over the base $B$ and which
we will
denote by $W$. In addition, note that the discriminant defined in 
equation~\eqref{eq:31A} is a section of the line bundle
\begin{equation}
\Delta \sim \cL^{12}
\label{eq:32A}
\end{equation}
over the base $B$. The zero locus of the discriminant section specifies a 
divisor of $B$, the discriminant curve. An elliptic fiber in $W$ is degenerate
if and only if it lies over a point in the discriminant curve.

Let us now demand that $W$ be a
Calabi--Yau threefold. It follows that we must require $c_1(TW)=0$. 
It can be shown that this, in turn, implies
\begin{equation}
\cL=K_B^{-1}
\label{eq:33}
\end{equation}
where $K_B$ is the canonical bundle on the base, $B$. 
Condition~\eqref{eq:33} is rather strong and, as stated earlier, restricts 
the allowed base spaces of an elliptically
fibered Calabi--Yau threefold to be rational (del Pezzo, Hirzebruch, as
well as certain blow--ups of Hirzebruch surfaces) and Enriques
surfaces (see for example \cite{na, g}). Henceforth,
we will only discuss Weierstrass fibrations that are, in addition, Calabi--Yau
threefolds. Using the fact that
$K_{B}^{-1}={\cal{O}}_{B}(c_{1}(B))$, it follows from $g_{2}$ and $g_{3}$
are sections
of the line bundles 
\begin{equation}
g_{2} \sim {\cal{O}}_{B}(4c_{1}(B)), \qquad  g_{3} \sim
{\cal{O}}_{B}(6c_{1}(B))
\label{eq:35} 
\end{equation}
This places constraints on the sections $g_{2}$ and $g_{3}$ that we
will return to below. In addition, note from ~\eqref{eq:32A}
and~\eqref{eq:33} that the discriminant is a section of the line bundle
\begin{equation}
\Delta \sim {\cal{O}}_{B}(12c_{1}(B))
\label{eq:36}
\end{equation}
This condition implies that the discriminant curve in $B$ must lie in the
class $12c_{1}(B)$, a strong restriction on allowed discriminant curves.

Generally, Weierstrass fibrations can be singular.
These singularities, however, can be removed by ``blowing--up'' the singular
points, producing a smooth elliptically fibered threefold that we will denote by
$X$. Despite the fact that $c_{1}(TW)=0$, the first Chern class of the
blown--up
fibration $X$ need not vanish and, hence, $X$ need not be a Calabi--Yau
threefold. It is possible to enforce the condition $c_{1}(TX)=0$, but only
at the cost of putting additional constraints on the sections $g_{2}$ and 
$g_{3}$. In this paper, we will always demand that $X$ is a Calabi--Yau
threefold and, hence, that $g_{2}$ and $g_{3}$ are suitably constrained.
 In this paper, for simplicity, we will
restrict the discussion to blow--ups such that each fiber of the induced
fibration $X \to B$ will still be a complex curve. This then 
places further constraints  on the allowed sections $g_{2}$ and $g_{3}$.
Finally, and conversely, it can be shown that any smooth elliptically fibered
Calabi--Yau threefold $X$ with a zero section can be obtained as a 
resolution of the singularities of a Weierstrass fibration.

To make these statements concrete, we now explicitly compute the discriminant
curve for an elliptically fibered Calabi--Yau threefold with base 
$B={\mathbb F}_{3}$.

\subsection*{Example 1:  $B={\mathbb F}_{3}$}

Consider a smooth elliptically fibered Calabi--Yau threefold $X$ with zero
section
$\sigma$ and base $B={\mathbb F}_{3}$. As discussed previously, 
there are two effective classes $\cE$ and $\cS$ that together form a basis 
of $H_{2}({\mathbb F}_{3}, \mathbb Z)$ with intersection numbers
\begin{equation}
\cE \cdot \cE= 0, \qquad  \cS \cdot \cS=-3, \qquad  \cE \cdot \cS= 1
\label{eq:37}
\end{equation}
The first and second Chern classes of $B={\mathbb F}_{3}$ are given by
\begin{equation}
c_{1}({\mathbb F}_{3})=2{\cal{S}}+5{\cal{E}}, \qquad 
c_{2}({\mathbb F}_{3})=4
\label{eq:37A}
\end{equation}
$X$ is the resolution of a Weierstrass fibration over the base $B={\mathbb
F}_{3}$ with sections $g_{2}$, $g_{3}$ satisfying~\eqref{eq:35} and
discriminant $\Delta$ defined by~\eqref{eq:31A} and satisfying~\eqref{eq:36}.
Let us first consider the consequences of~\eqref{eq:35}. To do this, we must
discuss several important assertions. The first of these, Claim I, is the
following.

\begin{itemize}

\item The zero locus of any section $g_{2} \sim {\cal{O}}_{B}(4c_{1}({\mathbb
F}_{3}))$ and $g_{3} \sim {\cal{O}}_{B}(6c_{1}({\mathbb F}_{3}))$ must
necessarily 
vanish along $\cS$ with order at least 2. 

\end{itemize}
The proof of this assertion is outlined as follows. First consider $g_{2}
\sim 
{\cal{O}}_{B}(4c_{1}({\mathbb F}_{3}))$. Note that one can always write
\begin{equation}
{\cal{O}}_{B}(4c_{1}({\mathbb F}_{3}))={\cal{O}}_{B}(2\cS) \otimes
{\cal{O}}_{B}(4c_{1}({\mathbb F}_{3})-2\cS)
\label{eq:37B}
\end{equation}
To prove Claim I, we need to show that any section 
$g_{2} \sim {\cal{O}}_{B}(4c_{1}({\mathbb F}_{3}))$ can be written in the form
\begin{equation}
g_{2}= s_{2\cS} \otimes G_{2}
\label{eq:37C}
\end{equation}
where
\begin{equation}
s_{2\cS} \sim {\cal{O}}_{B}(2\cS), \qquad 
G_{2} \sim {\cal{O}}_{B}(4c_{1}({\mathbb F}_{3})-2\cS)
\label{eq:37D}
\end{equation}
Denote the vector space of all sections of 
${\cal{O}}_{B}(4c_{1}({\mathbb F}_{3}))$ and ${\cal{O}}_{B}(4c_{1}({\mathbb
F}_{3})-2\cS)$ by $H^{0}({\mathbb F}_{3}, 4c_{1}({\mathbb F}_{3}))$ and 
$H^{0}({\mathbb F}_{3}, 4c_{1}({\mathbb F}_{3})-2\cS)$ respectively. The
decomposition~\eqref{eq:37C} will be satisfied if the  dimensions of 
these two vector spaces are identical, which we now proceed to show. 
To do this, consider the exact sequence
\begin{equation}
0 \to H^0({\mathbb F}_{3}, 4 c_1({\mathbb F}_{3}) - \cS) \to H^0({\mathbb
F}_{3},
4 c_1({\mathbb F}_{3})) \to H^0(\cS, 4 c_1({\mathbb F}_{3})|_{\cS})
\label{eq:38}
\end{equation}
Note, however, that
\begin{equation}
H^0(\cS, (4 c_1({\mathbb F}_{3}))|_{\cS})= H^0({\mathbb {CP}}^{1},  
4c_1({\mathbb F}_{3}) \cdot \cS)=  H^0({\mathbb {CP}}^{1}, -4)=0
\label{eq:39}
\end{equation}
where we have used~\eqref{eq:37} and~\eqref{eq:37A}. Hence, 
dim$H^0({\mathbb F}_{3}, 4 c_1({\mathbb F}_{3}) - \cS)=$ 
dim$H^0({\mathbb F}_{3}, 4 c_1({\mathbb F}_{3}))$.
Similarly, consider the exact sequence
\begin{equation}
0 \to H^0({\mathbb F}_{3}, 4 c_1({\mathbb F}_{3}) - 2\cS) \to H^0({\mathbb
F}_{3},
4 c_1({\mathbb F}_{3})-\cS) \to H^0(\cS, (4 c_1({\mathbb F}_{3})-\cS)|_{\cS})
\label{eq:40}
\end{equation}
and the relation
\begin{equation}
H^0(\cS, (4 c_1({\mathbb F}_{3})-\cS)|_{\cS})= H^0({\mathbb {CP}}^{1},  
(4c_1({\mathbb F}_{3})-\cS) \cdot \cS)=  H^0({\mathbb {CP}}^{1}, -1)=0
\label{eq:41}
\end{equation}
Therefore, dim$H^0({\mathbb F}_{3}, 4 c_1({\mathbb F}_{3}) - 2\cS)=$
dim$H^0({\mathbb F}_{3}, 4 c_1({\mathbb F}_{3}) - \cS)$. Combining with the
above relation implies 
\begin{equation}
dimH^0({\mathbb F}_{3}, 4 c_1({\mathbb F}_{3}) - 2\cS)=
dimH^0({\mathbb F}_{3}, 4 c_1({\mathbb F}_{3}))
\label{eq:43}
\end{equation}
This establishes the claim for the section $g_{2} \sim {\cal{O}}_{B}(4c_{1}
({\mathbb F}_{3}))$. The proof is similar for the section
$g_{3} \sim {\cal{O}}_{B}(6c_{1}({\mathbb F}_{3}))$ and establishes that
\begin{equation}
g_{3}= s_{2\cS} \otimes G_{3}
\label{eq:43A}
\end{equation}
where
\begin{equation}
s_{2\cS} \sim {\cal{O}}_{B}(2\cS), \qquad 
G_{3} \sim {\cal{O}}_{B}(6c_{1}({\mathbb F}_{3})-2\cS)
\label{eq:43B}
\end{equation}

Simply put, Claim I says that if in local coordinates u,v on $B={\mathbb
F}_{3}$ the curve $\cS$ is given by $u=0$, then, near $\cS$, $g_{2}$ and 
$g_{3}$ must be of the form
\begin{equation}
g_{2}=u^{2}G_{2}, \qquad  g_{3}=u^{2}G_{3}
\label{eq:44}
\end{equation}
However, it does not specify the form of $G_{2}$ and $G_{3}$, which may or
may 
not vanish on $\cS$. The properties of $G_{2}$ and $G_{3}$ are further refined
in a second claim, Claim II, which we state without proof.

\begin{itemize}

\item The divisors $4 c_1({\mathbb F}_{3}) - 2\cS$ and $6 c_1({\mathbb F}_{3}) -
2\cS$ are ``very ample''.

\end{itemize}
For a divisor $D$ in a space $B$ to be very ample means that a) for any
point $p
\in B$ there must be a curve in the class of $D$ that passes through $p$ and
b) for any two disjoint points $p,q \in B$ there must be a curve in the class
of $D$ that passes through $p$ but not $q$, and a curve that passes through
$q$ but not $p$. Claim II  tells us that there must exist sections 
$g_{2}$ and $g_{3}$ that vanish at exactly order 2 along $\cS$, that is, for
which $G_{2}(0,v)$ and $G_{3}(0,v)$ are non-vanishing.

Using these results, we now turn to the discriminant curve $\Delta$ defined
in~\eqref{eq:31A} and the consequences of~\eqref{eq:36}. The third assertion,
Claim $III$, is the following.

\begin{itemize}

\item The zero locus of the discriminant $\Delta= g_{2}^{3}-27g_{3}^{2} \sim
\cO_{B}(12c_{1}({\mathbb F}_{3})   )$ must necessarily vanish along $\cS$
with 
order exactly 4. Furthermore, the zero locus of $\Delta$ also vanishes 
along the curve described by $\Sigma=20\cS+60\cE$. The curves $\cS$ and 
$\Sigma$ do not intersect, that is, $\cS \cdot \Sigma =0$.

\end{itemize} 

The proof of Claim III is as follows. First note from the definition of the
discriminant section $\Delta= g_{2}^{3}-27g_{3}^{2}$ and~\eqref{eq:37C},
~\eqref{eq:43A} that
\begin{equation}
\Delta=s_{2\cS}^{2} \otimes \Delta_{12c_{1}(B)-4\cS}
\label{eq:45}
\end{equation}
where
\begin{equation}
\Delta_{12c_{1}(B)-4\cS}=s_{2\cS} \otimes G_{2}^{3} -27G_{3}^{2}
\label{eq:46}
\end{equation}
>From Claim I we know that
the zero locus of $\Delta$ vanishes along $\cS$ with order at least 4.  A in 
Claim I, we can show that the zero
locus of $\Delta$ vanishes along $\cS$ with order exactly 4. Simply put, this
implies, in terms of the local coordinates u,v on $B={\mathbb F}_{3}$,
that near $\cS$ the discriminant $\Delta$ must be of the form
\begin{equation}
\Delta= u^{4}(-27G_{3}(0,v)^{3})
\label{eq:46A}
\end{equation}
where the factor $-27G_{3}(0,v)^{3}$ is non-vanishing. Now, note
from~\eqref{eq:37A} that
\begin{equation}
12c_{1}({\mathbb F}_{3})-4\cS=20\cS +60\cE
\label{eq:47}
\end{equation} 
We conclude that $\Delta$ must also vanish along a curve of the form $\Sigma=
20\cS +60\cE$. Finally, using equation~\eqref{eq:37} it is easy to see that 
$\cS \cdot \Sigma= 0$. This establishes Claim III. Finally, we prove a fourth
assertion, Claim IV.

\begin{itemize}

\item $\cS$ is a smooth rational curve in $B={\mathbb F}_{3}$. A torus over any
point
on the curve $\cS$ degenerates as Kodaira type $IV$. The curve $\Sigma$ in
$B={\mathbb F}_{3}$ is
smooth except at 200 cusps. The fibers degenerate as Kodaira type $I_{1}$ over
the smooth parts of the curve and as Kodaira type $II$ over each of the
cusps.

\end{itemize}
To prove Claim IV,  first consider the curve $\cS$. It follows
from~\eqref{eq:44}, Claim II and~\eqref{eq:46A} that, in local coordinates
near
$\cS$, sections $g_{2}$, $g_{3}$ and $\Delta$ have the form
\begin{equation}
g_{2}=u^{2}G_{2}, \qquad  g_{3}=u^{2}G_{3}, \qquad 
\Delta= u^{4}(-27G_{3}(0,v)^{3})
\label{eq:48}
\end{equation} 
Using Table 1, we find that any torus over a point on the curve $\cS$ 
degenerates as a Kodaira fiber of Type $IV$. Now consider the curve
specified by
$\Sigma$. There are two ways in which the section 
$\Delta_{12c_{1}(B)-4\cS}=s_{2\cS} \otimes G_{2}^{3} -27G_{3}^{2}$ can vanish
on $\Sigma$. First, $G_{2}$ and $G_{3}$ can vanish simultaneously. We see
from 
~\eqref{eq:37D} and~\eqref{eq:43B} that this will occur precisely at the
points of intersection of the divisors $4 c_1({\mathbb F}_{3}) - 2\cS$ 
and $6 c_1({\mathbb F}_{3})-2\cS$. Since these divisors, by Claim II, are very
ample, then, $G_{2}$ and $G_{3}$ can be chosen so that,
near each such intersection point,
they are local coordinates of
the base. Denoting $G_{2}=s$ and $G_{3}=t$, then the
discriminant curve is described by the equation
\begin{equation}
f(s,t)s^{3}-27t^{2}=0
\label{eq:51}
\end{equation} 
where the function $f(s,t)$, which represents $u^{2}$ evaluated in $s,t$
coordinates, is non-vanishing in the neighborhood of the curve. One
can check that this curve has a cusp at the point $(s,t)=(0,0)$. That is, each
intersection point corresponds to a cusp of the curve $\Sigma$. It follows
that in the neighborhood of a cusp, the Weierstrass equation~\eqref{eq:31}
for the elliptic fiber can be written as
\begin{equation}
y^{2}=4x^{3}-f(s,t)sx-f(s,t)t
\label{eq:52}
\end{equation} 
where we have used the affine coordinate $z=1$. This 
defines a smooth threefold. Exactly at the cusp, this
equation becomes
\begin{equation}
y^{2}=4x^{3}
\label{eq:53}
\end{equation} 
which is well known to describe a fiber of Kodaira type $II$.
Clearly, the total number of such cusps,
$N_{cusps}$, of the curve $\Sigma$ is given by
\begin{equation}
N_{cusps} = (4 c_1({\mathbb F}_{3}) - 2\cS) \cdot 
(6 c_1({\mathbb F}_{3}) -2\cS) = 200
\label{eq:50}
\end{equation} 
where we have used expressions~\eqref{eq:37} and~\eqref{eq:37A}. Away from
these 200 cusps, the curve $\Sigma$ is smooth. Along
the smooth part of the curve, we can introduce new local coordinates $S=
f(s,t)s^{3}-27t^{2}$ and an independent variable $T$ such that $\Sigma$ is
defined by $S=0$. We see then that the
sections $g_{2}$, $g_{3}$ and $\Delta$, in a neighborhood of the smooth part
of the curve $\Sigma$, have the form
\begin{equation}
g_{2}=S^{0}{\cal{G}}_{2}, \qquad g_{3}=S^{0}{\cal{G}}_{3}, \qquad
\Delta=S^{1}f^{2}
\label{eq:54}
\end{equation} 
where ${\cal{G}}_{2}, {\cal{G}}_{2}$ and $f$ are functions of $S$ and $T$
which 
do not vanish on $\Sigma$.
It follows from Table 1 that the torus over any point on the smooth part of
the curve $\Sigma$ degenerates as Kodaira type $I_{1}$. This completes the
proof of Claim IV.

\vspace{30pt}

\begin{tabular}{|l|c|c|c|} 
\hline
Kodaira type & A-D-E & monodromy & N,L,K \\
\hline
$I_n$ & $A_{n-1}$ & $\begin{pmatrix} 1 &  n \\ 0 &  1 \end{pmatrix} $ &
$N=n, L=0, K=0$ \\ 
\hline
$II$ & & $\begin{pmatrix}1 & 1\\ -1 & 0 \end{pmatrix}$ & ${N=2, L>0, K=1}$ \\
\hline
$III$ & $A_1$ & $\begin{pmatrix}0 & 1\\ -1 & 0\end{pmatrix}$ & $N=3, L=1,
K>1$\\
\hline
$IV$ & $A_2$ & $\begin{pmatrix}0 & 1\\ -1 & -1 \end{pmatrix}$ & $N=4, L>1,
K=2$\\
\hline
$I_0^*$ & $D_4$ & $\begin{pmatrix} -1 & 0\\ 0 & -1 \end{pmatrix}$ & $N=6,
L>1, K>2$\\
\hline
$I_n^*$ & $D_{n+4}$ & $\begin{pmatrix}-1 & -n \\ 0 & -1\end{pmatrix}$ & $N
= 6+n, L=2, K=3$\\
\hline
$IV^*$ & $E_6$ & $\begin{pmatrix} -1 & -1 \\ 1 & 0 \end{pmatrix}$ & $N = 8,
L>2, K=4$\\
\hline
$III^*$ & $E_7$ & $\begin{pmatrix} -0 & -1 \\ 1 & 0\end{pmatrix}$ & $N = 9,
L=3, K>4$\\
\hline
$II^*$ & $E_8$ & $\begin{pmatrix} 0 & -1 \\ 1 & 1 \end{pmatrix}$ & $N = 10,
L>3, K=4$\\
\hline
\end{tabular}

\vspace{30pt}

{\small \noindent
Table 1: The integers $N$, $L$ and $K$ characterize the behavior
of $\Delta$, $g_{2}$ and $g_{3}$ near the discriminant locus $u=0$; 
$\Delta = u^Na , g_{2} = u^Lb$ , and 
$g_{3} =u^Kc$ .}

\vspace{30pt}

To conclude, the discriminant curve for an elliptically fibered
Calabi--Yau threefold with base $B={\mathbb F}_{3}$ has the following
structure. First,
\begin{equation}
\Delta=\cS \cup \Sigma
\label{eq:55}
\end{equation} 
where
\begin{equation}
\Sigma=20\cS+60\cE
\label{eq:55A}
\end{equation} 
and
\begin{equation}
\cS \cdot \Sigma=0
\label{eq:56}
\end{equation} 
The component curve $\cS$ is smooth, whereas the other component curve
$\Sigma$ is smooth except at 200 points, where it degenerates into cusps. 
We first list the Kodaira type for the fibers over the smooth parts of 
these curves.

\begin{itemize} 

\item $\cS$-- Kodaira type $IV$

\item $\Sigma$-- Kodaira type $I_{1}$

\end{itemize}

The Kodaira type over the cusp points of $\Sigma$ are

\begin{itemize}

\item $\Sigma$-- cusps-- Kodaira type $II$

\end{itemize}

This is shown pictorially in Figure $1$.

\onefigure{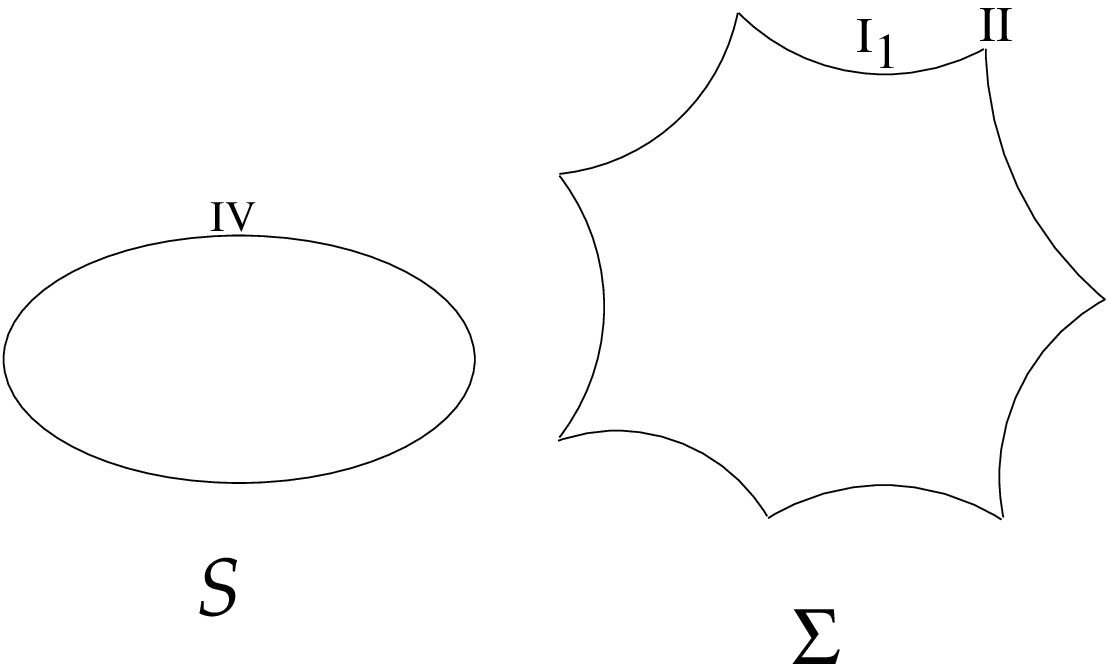}{Schematic illustration of the discriminant curve in
Example $1$. The fibers over the $\cS$ component are of Kodaira type $IV$,
while 
over the smooth part of the $\Sigma$ component they are of 
Kodaira type $I_{1}$.  The fibers over the cusps of $\Sigma$ are Kodaira
type $II$.}

We now apply this technology to the case of interest in this paper, an
elliptically fibered Calabi--Yau threefold with the base $B=\hat{\mathbb
F}_{3}$ introduced above. Since the computations are similar to the ones just
described, we will only present the results. (These examples were
first constructed torically in \cite{ar}).

\subsection*{Example 2: $B=\hat{\mathbb F}_{3}$}

Consider a smooth elliptically fibered Calabi--Yau threefold $X$ with zero
section
$\sigma$ and base $B=\hat{\mathbb F}_{3}$. As discussed previously, 
there are three effective classes $\cS$, $\cF$ and $\cG$ that together form
a basis 
of $H_{2}(\hat{\mathbb F}_{3}, \mathbb Z)$ with intersection numbers
\begin{equation}
{\cal{S}} \cdot {\cal{S}}= -3,  \qquad  {\cal{F}} \cdot {\cal{F}}=-1,
\qquad {\cal{G}} \cdot {\cal{G}}= -1
\label{eq:57}
\end{equation}
and
\begin{equation}
{\cal{S}} \cdot {\cal{F}}= 1,  \qquad  {\cal{S}} \cdot {\cal{G}}=0,
\qquad {\cal{G}} \cdot {\cal{F}}= 1
\label{eq:58}
\end{equation}
The first and second Chern classes of $B=\hat{\mathbb F}_{3}$ are given by
\begin{equation}
c_{1}(\hat{\mathbb F}_{3})=2{\cal{S}}+5{\cal{F}}+4{\cal{G}}, \qquad 
c_{2}(\hat{\mathbb F}_{3})=5
\label{eq:59}
\end{equation}
It is helpful in the following to define the class
\begin{equation}
\sigma= \cS+3\cF+2{\cal{G}}
\label{eq:60}
\end{equation}
$X$ is the resolution of a Weierstrass fibration over the base $B=\hat{\mathbb
F}_{3}$ with sections $g_{2}$, $g_{3}$ satisfying~\eqref{eq:35} and
discriminant $\Delta$ defined by~\eqref{eq:31A} and satisfying~\eqref{eq:36}.
Using this data, one can, as outlined above, compute the discriminant curves.
We will simply state the results.

\subsection*{Case 1:}

The discriminant curve is composed of three components
\begin{equation}
\Delta= \cS \cup \sigma \cup \Sigma
\label{eq:60A}
\end{equation}
where
\begin{equation}
\Sigma=16\cS+ 54\cF+ 44{\cal{G}}
\label{eq:60B}
\end{equation}
It follows from~\eqref{eq:57} and~\eqref{eq:58} that
\begin{equation}
\cS \cdot \sigma =0, \qquad \sigma \cdot \Sigma=44, \qquad \cS \cdot \Sigma=6
\label{eq:60C}
\end{equation}
The intersection number $\sigma \cdot \Sigma=44$ corresponds to 36
intersection points of $\sigma$ with $\Sigma$, 28 of which
are simple normal crossing, whereas 8 have tangential intersections
of multiplicity 2. The 6 intersection points of $\cS$ with $\Sigma$ are all
simple normal crossings. The number of cusps of the component curve $\Sigma$
(see \cite{GM}) is
\begin{equation}
N_{cusps}=150
\label{eq:60D}
\end{equation}
Away from the intersection points and the cusps, the discriminant curve
is smooth. The Kodaira types for the smooth parts of the component curves
$\sigma$ and $\Sigma$ are given by

\begin{itemize}

\item $\sigma$-- Kodaira type $I_{2}$

\item $\Sigma$-- Kodaira type $I_{1}$

\item $\cS$-- Kodaira type $I_{0}^{*}$

\end{itemize}
The Kodaira type of the $\sigma \cdot \Sigma$ and $\cS \cdot \Sigma$
intersection points are

\begin{itemize}

\item $\sigma \cdot \Sigma$-- 28 simple normal crossings-- Kodaira type
$I_{3}$

\item $\sigma \cdot \Sigma$-- 8 tangential crossings-- Kodaira type $III$

\item $\cS \cdot \Sigma$-- 6 simple normal crossings-- Not Kodaira 

\end{itemize}
Finally, over each of the cusp points of curve $\Sigma$ we find

\begin{itemize}

\item $\Sigma$-- cusps-- Kodaira type $II$

\end{itemize}
This is shown pictorially in Figure 2.

\onefigure{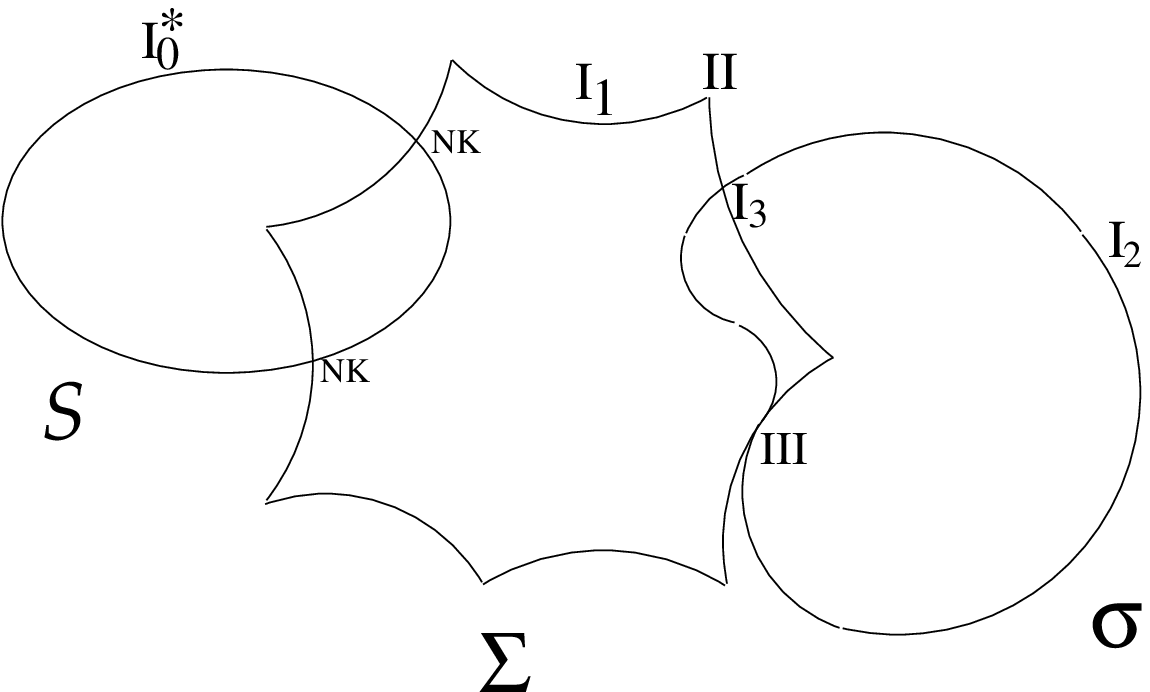}{Schematic illustration of the discriminant curve for 
Case $1$ of Example $2$.  
The fibers over the smooth parts of the curves $\cS$, $\sigma$ and $\Sigma$
are of Kodaira type $I_{0}^{*}$, $I_{2}$ and $I_{1}$ respectively. The
fibers over
the cusp points of $\Sigma$ are of Kodaira type $II$.
There are non-Kodaira fibers (NK) over the points where the
$\cS$ component intersects the $\Sigma$ components normally. There are
$I_3$ fibers where the $\sigma$ component meets the $\Sigma$ component
normally, and fibers of type $III$ where the $\sigma$ component
intersects the $\Sigma$ component tangentially.}

We now present three more discriminant curves.
In all of these, the discriminant is composed of the
three component curves 
\begin{equation}
\Delta= \cS \cup \sigma \cup \Sigma
\label{eq:61}
\end{equation}
where
\begin{equation}
\cS \cdot \sigma =0, \qquad \sigma \cdot \Sigma=44
\label{eq:61A}
\end{equation}
The intersection number $\sigma \cdot \Sigma=44$ corresponds to 36
intersection points of $\sigma$ with $\Sigma$, 28 of which
correspond to simple normal crossing whereas 8 have tangential intersections
of multiplicity 2. The component curve $\Sigma$ always has a finite number of
cusps. Away from the intersection points and the cusps, the discriminant curve
is smooth. The Kodaira type for the smooth parts of the component curves
$\sigma$ and $\Sigma$ are in all cases given by

\begin{itemize}

\item $\sigma$-- Kodaira type $I_{2}$

\item $\Sigma$-- Kodaira type $I_{1}$

\end{itemize}

Similarly, in all cases the Kodaira type of the $\sigma \cdot \Sigma$
intersection points are

\begin{itemize}

\item $\sigma \cdot \Sigma$-- 28 simple normal crossings-- Kodaira type
$I_{3}$

\item $\sigma \cdot \Sigma$-- 8 tangential crossings-- Kodaira type $III$

\end{itemize}

The final shared characteristic is that over the cusp points
of the curve $\Sigma$ we always have

\begin{itemize}

\item $\Sigma$-- cusps-- Kodaira type $II$

\end{itemize}

\subsection*{Case 2:}

The component curve $\Sigma$ is given by
\begin{equation}
\Sigma=14\cS+54\cF+44{\cal{G}}
\label{eq:64}
\end{equation}
with intersection number 
\begin{equation}
\cS \cdot \Sigma= 12
\label{eq:64AA}
\end{equation}
This corresponds to 6 intersection points of $\cS$ with $\Sigma$, 
each a tangential crossing with multiplicity 2. 
In addition, the number of cusps of the curve $\Sigma$ is
\begin{equation}
N_{cusps}=142
\label{eq:64A}
\end{equation}
Over the smooth parts of component curve $\cS$ we find

\begin{itemize} 

\item $\cS$-- Kodaira type $IV^{*}$

\end{itemize}
The Kodaira type over the remaining intersection points are

\begin{itemize}

\item $\cS \cdot \Sigma$-- 6 tangential crossing-- Not Kodaira 

\end{itemize}

\subsection*{Case 3:}

The component curve $\Sigma$ is given by
\begin{equation}
\Sigma=18\cS+54\cF+44{\cal{G}}
\label{eq:66AA}
\end{equation}
with intersection number given by
\begin{equation}
\cS \cdot \Sigma= 0
\label{eq:66}
\end{equation}
In addition, the number of cusps of the curve $\Sigma$ is
\begin{equation}
N_{cusps}=152
\label{eq:66A}
\end{equation}
The Kodaira type for the fibers over the smooth part of the component curve
$\cS$ is

\begin{itemize} 

\item $\cS$-- Kodaira type $IV$

\end{itemize}

\subsection*{Case 4:}

The component curve $\Sigma$ is given by
\begin{equation}
\Sigma= 13\cS+54\cF+44{\cal{G}}
\label{eq:68AA}
\end{equation}
The remaining intersection number is now given by
\begin{equation}
\cS \cdot \Sigma= 15
\label{eq:68}
\end{equation}
The intersection number $\cS \cdot \Sigma=15$ corresponds to 5
intersection points of $\cS$ with $\Sigma$, all of which  are tangential 
crossings of multiplicity 3.
In addition, the number of cusps of the curve $\Sigma$ is 
\begin{equation}
N_{cusps}=137
\label{eq:68A}
\end{equation}
The Kodaira type for the fibers over the smooth part of $\cS$ is

\begin{itemize} 

\item $\cS$-- Kodaira type $III^{*}$

\end{itemize}
The Kodaira type over the remaining intersection points are

\begin{itemize}

\item $\cS \cdot \Sigma$-- 5 tangential crossings-- Kodaira type $III^{*}$

\end{itemize}

To conclude, we have presented a general formalism for constructing the
discriminant curves of elliptically fibered Calabi--Yau threefolds. This
formalism was applied to a specific Calabi--Yau threefold with a
blown--up Hirzebruch base $B=\hat{\mathbb F}_{3}$. Each of the discriminant
curves was shown to be composed of smooth sections, as well as cusps,
simple normal
crossing points and tangential crossings. In this paper, we will be concerned
with five--branes wrapped on elliptic fibers near
the smooth parts of the discriminant curves only. We leave the discussion
of five--branes near the cusps, simple normal crossings and
tangential crossing points to another publication.


\section{Monodromy, Massless States and Brane Junctions:}


In this section, we will compute the massless spectrum on the worldvolume of
five--branes wrapped on elliptic fibers near the smooth parts of discriminant
curves. This is accomplished by demonstrating the equivalence, in this
context, of degenerating M membranes to string junctions, and using the
junction lattice techniques developed in \cite{bartonone,bartontwo}.
For concreteness, we present our results within the context of
the three--family, $SU(5)$ model compactified on a
Calabi--Yau threefold with blown--up Hirzebruch base $B=\hat{\mathbb F}_{3}$ 
presented above. This formalism, however, is applicable to any theory
compactified on elliptically fibered Calabi--Yau threefolds.

We begin by briefly reviewing the worldvolume theory of an M--theory 
five--brane wrapped on a smooth elliptic fiber $\cC_{2}$ located far from the 
discriminant locus. First, consider a five--brane in flat space--time. The
massless 
worldvolume degrees of freedom then form a six--dimensional $(2,0)$
supersymmetric
tensor multiplet. The bosonic content of this multiplet consists of five
scalar
fields, $X^{i}$, $i=1,..,5$ describing translations in the transverse
directions together with a two--form $\cB$ whose field strength $H=d\cB$ is
self--dual, $H=*H$. When the five--brane is wrapped on a smooth elliptic
fiber of a Calabi--Yau threefold, four of the scalars become moduli
for the location of the fiber, that is, they are coordinates of the
base surface $B$. We denote these four fields by $u_{1}, u_{2}, v_{1}$, and
$v_{2}$. The fifth scalar, which we denote by $y$, continues to 
parameterize translations in
the remaining transverse direction. The behavior of the self--dual field
strength is a little more complicated. The cohomology classes of a smooth
elliptic
fiber are those of a real two--torus, that is, $H^{0}$, $H^{1}$ and $H^{2}$
with
dimensions one, two and one respectively. We denote a basis of $H^{0}$ by $1$,
of $H^{1}$ by the two harmonic one--forms $\lambda_{c}$, $c=1,2$ and of 
$H^{2}$ by the harmonic volume form $\Omega$. If we decompose the five--brane
worldvolume as $\cC_{2} \times M_{4}$, then the field strength $H$ 
can be expanded as
\begin{equation}
H=da \wedge \Omega+ F^{c}\wedge \lambda_{c}+ h 
\label{eq:69}
\end{equation}
where the four--dimensional fields are a scalar ``axion'' $a$, two $U(1)$
vector
fields $F^{c}=dA^{c}$ and a three--form field strength $h=db$. However, not
all of these fields are independent because of the self--duality condition
$H=*H$. Applying this condition leads to the constraints
\begin{equation}
F^{2}=*F^{1}, \qquad h=*da
\label{eq:70}
\end{equation}
We conclude that the two--form $\cB$ decomposes into an axionic scalar field
$a$ and a single $U(1)$ gauge connection $A_{\mu}$. Putting everything
together, we
find  that for a five--brane compactified on a smooth elliptic fiber
$\cC_{2}$,
the bosonic worldvolume fields on $M_{4}$ are $A_{\mu}, a, y, u_{1}, u_{2}, 
v_{1}, v_{2}$. These fields, along with their fermionic superpartners, form
an $N=4$ vector supermultiplet. The low energy worldvolume field theory on
$M_{4}$ exhibits $N=4$ supersymmetry, but this is broken to $N=1$
supersymmetry by higher dimension operators. The $N=4$ vector
supermultiplet decomposes under $N=1$ supersymmetry into an 
Abelian Yang--Mills multiplet
with bosonic field $A_{\mu}$, and three chiral multiplets with bosonic
fields $Y=y+ia$, $u=u_{1}+iu_{2}$ and $v=v_{1}+iv_{2}$ respectively. Note
that none of these  chiral supermultiplets is charged. Before continuing, we
record the fact that the first homology group of a smooth elliptic fiber, 
$H_{1}(\cC_{2}, {\mathbb Z})$, has dimension two and a basis of one--cycles 
$\omega_{c}$, $c=1,2$. It follows that, in this basis, any cycle in 
$H_{1}(\cC_{2}, {\mathbb Z})$ is specified by $(p,q)$, where $p,q \in {\mathbb
Z}$.

We see from the examples given in the previous section that elliptic fibers
above different points on the smooth parts of discriminant curves can have
very different degeneration characteristics.  The simplest, which, for
example,
occurs at the smooth parts of the $\Sigma$ component curve,  
is associated with Kodaira type $I_{1}$. We will discuss this case
first.

\subsection*{Kodaira Type $I_{1}$:}

In the neighborhood of a 
smooth point on the discriminant curve, one can always define two special
complex coordinates, a coordinate $u$ transverse to the curve and a
coordinate $v$ along the curve. Note that previously these coordinates were
designated by various symbols, such as $S$ and $T$, but, henceforth, 
we will call the complex coordinates near any component of 
the discriminant curve $u$ and $v$. Now pick an arbitrary point on the smooth
part of the discriminant curve with a fiber of Kodaira type $I_{1}$, and
choose the origin of the $u,v$ coordinates to be at that point. 
Recall from~\eqref{eq:54} that, in
the neighborhood of this point, the associated sections have the form
\begin{equation}
g_{2}=u^{0}{\cal{G}}_{2}, \qquad g_{3}=u^{0}{\cal{G}}_{3}, \qquad
\Delta= u^{1}f^{2}
\label{eq:71}
\end{equation}
where ${\cal{G}}_{2}$, ${\cal{G}}_{3}$ and $f$, to leading order in $u$, 
are non--zero functions of $v$ only. We
see from~\eqref{eq:31A} that the first two functions must satisfy
\begin{equation}
{\cal{G}}_{2}^{3}-27{\cal{G}}_{3}^{2}=0               
\label{eq:72}
\end{equation}
It follows from~\eqref{eq:31} and~\eqref{eq:71} that, at the origin, the
Weierstrass form of the elliptic fiber becomes
\begin{equation}
y^{2}=4x^{3}-\cG_{2}x-\cG_{3}
\label{eq:73}
\end{equation}
where we have used affine coordinates with $z=1$. This is the standard
Weierstrass representation of an elliptic fiber of Kodaira type $I_{1}$.

In the neighborhood of a point on the smooth part of the 
discriminant curve, the symmetry between the complex moduli $u$ and $v$, 
that exists far from the discriminant locus, is broken. 
Hence, near such a
point, the $N=4$ supersymmetry of the low energy $M_{4}$ worldvolume theory of
the wrapped five--brane is reduced to $N=2$ supersymmetry. 
The $N=4$ vector supermultiplet discussed above then decomposes into two $N=2$
supermultiplets, an Abelian Yang-Mills multiplet with bosonic field content 
$A_{\mu}$, $u$ and a hypermultiplet with the bosonic fields $Y$, $v$. As is
the
case far from the discriminant curve, the supersymmetry is broken to $N=1$
supersymmetry by higher dimension operators.

To discuss the monodromy, we must restrict to the fibration over a curve that
intersects the discriminant locus at the origin. 
 The ``generic'' curves are transverse to the
discriminant locus, that is, their intersection multiplicity  
with the discriminant locus is $1$. Surfaces over curves
with multiplicity greater than $1$ are singular at the discriminant point,
while those over generic curves are smooth. 
 Therefore, we
will choose the intersection surface to be the fibration over a generic curve.
Note that the path $\cP$ defined by $v=0$, which has coordinate $u$, is
tranverse to the discriminant curve and, hence, is generic. 
For specificity, we will take the intersecting surface to be the elliptic
fibration over $\cP$, which we denote by ${\cal{T}}$. Restricted to this surface, 
the discriminant curve appears
as a point in the one--dimensional complex base $\cP$. Furthermore, the
forms of $g_{2}$, $g_{3}$, and $\Delta$, as well as the Weierstrass form of the
elliptic fiber, restricted to this twofold remain those
given in~\eqref{eq:71} and~\eqref{eq:73} with $\cG_{2}$, $\cG_{3}$, 
and $f$ evaluated at $v=0$. It follows from Table $1$ that, within ${\cal{T}}$, 
the degeneration of the fiber near the discriminant
point remains Kodaira type $I_{1}$. Since the problem has 
now been reduced to an elliptic twofold over a one--dimensional base, we can apply
standard Kodaira theory to analyze the monodromy.

Before proceeding, it is important to discuss the structure of the 
elliptic fibration ${\cal{T}}$ over the generic base curve $\cP$. 
${\cal{T}}$ will be a $K3$ twofold
if and only if  $\cP \cdot \cP=0$, which guarantees $c_{1}({\cal{T}})=0$
and $\cP$ is rational, which implies that $H_{1}({\cal{T}}, {\mathbb
Z})=0$. For the
specific threefold base $B=\hat{\mathbb F}_{3}$ and the 
smooth parts of the discriminant
curves discussed in this paper, both of these properties are satisfied.
However, they need not be true for general threefold bases and
discriminant curves, and have to be checked on a case by case basis. In
general, ${\cal{T}}$ need not be a $K3$ surface.

The $SL(2, {\mathbb Z})$ monodromy transformation for a type $I_{1}$ fiber can be 
found from Table $1$, and is given by
\begin{equation}
A=\begin{pmatrix} 
                  1 & 1 \\
                  0 & 1
                  \end{pmatrix}
\label{eq:74}
\end{equation}
This transformation acts on the elements of $H_{1}(\cC_{2}, {\mathbb Z})$ and
has, up to multiplication by non--zero integers, a single eigenvector
\begin{equation}
\vec{{\cal{V}}}=(1,0)
\label{eq:75}
\end{equation}
with eigenvalue $+1$. The meaning of this result is the 
following. Consider an
elliptic fiber $\cC_{2}$ over a point, $z$, near, but not at, the discriminant
locus. 
Then the one--cycle, $\vec{{\cal{V}}}=(1,0)$, is the unique cycle (up to multiplication) in 
$H_{1}(\cC_{2}, {\mathbb Z})$
that contracts to zero as the base point of the fiber 
is moved to the discriminant. That
is, $\vec{{\cal{V}}}=(1,0)$ is the unique ``vanishing cycle''
(see for example \cite{BpvV}). 

The physical implications of this arise as follows. Fix a curve in the
base from point $z$ to the discriminant locus. Now consider a membrane in
${\cal{T}}$ bounded by $\vec{{\cal{V}}}=(1,0)$ in the elliptic fiber over $z$, whose
intersection with the fiber over each point on this curve is the vanishing
cycle. Note that this membrane ``ends'' on the degenerate cycle of the $I_{1}$
fiber over the discriminant locus. A ${\cal{T}}$--membrane of this type 
is shown in Figure $3$. One can show that membranes constructed in this manner
have representatives which are holomorphic two-cycles in ${\cal{T}}$, 
albeit with respect to a different
complex structure than the usual one compatible with the elliptic
fibration. The class of such membranes lies in 
the relative homology $H_{2}({\cal{T}}/F, {\mathbb Z})$ and 
will be denoted by $\vec{v}$.
If ${\cal{T}}=K3$, then  \cite{Yau}  the self--intersection number of
any class $\vec{v}$ containing a holomorphic two--cycle is given by
\begin{equation}
\vec{v} \cdot \vec{v}= 2g-2 +b
\label{eq:75A}
\end{equation}
where $g$ is the genus and $b$ is the number of boundaries. In our case, $g=0, b=1$
and, hence
\begin{equation}
\vec{v} \cdot \vec{v} =-1
\label{eq:76}
\end{equation}
When ${\cal{T}}$ is not a $K3$ surface, the situation requires further analysis.
However, we are able to show that for a general surface ${\cal{T}}$
self--intersection~\eqref{eq:76} continues to hold. This is done as follows. 
Consider a second membrane $\vec{v}'$ homologous to $\vec{v}$. These membranes
intersect only at the collapsed cycle in the Kodaira fiber 
over the discriminant point and at the $(1,0)$ cycle in the fiber 
$\cC_{2}$ over $z$. Using the fact that the self--intersection of the $(1,0)$
cycle is zero, it follows that the intersection number of $\vec{v}'$ with
$\vec{v}$ and, hence, the self--intersection of $\vec{v}$ is $\pm 1$. Since
this result is topological, we see from~\eqref{eq:76} that for any
surface ${\cal{T}}$, not just $K3$, $\vec{v} \cdot \vec{v}=-1$.

As long as the five--brane is wrapped on an elliptic fiber reasonably far from
the discriminant, the fluctuations of this membrane are massive and can be
ignored in the low energy effective theory. However, as the elliptic fiber
approaches the discriminant point, these fluctuations become less and less
heavy, finally becoming exactly massless when the five--brane is wrapped on
the degenerate $I_{1}$ Kodaira fiber. Therefore, at the discriminant locus we
expect light BPS states to enter the $M_{4}$ low energy theory of
the wrapped five--brane.                                           

To compute these states, first note that any cycle of the form $(p,0)$, where
$p \in {\mathbb Z}$, is an eigenvector of the monodromy $A$, 
not just $(1,0)$. The class of the associated membrane is denoted by 
\begin{equation}                                                              
\vec{J}=n \vec{v}                       
\label{eq:77}                                                               
\end{equation}
where $n \in {\mathbb Z}$. The vanishing cycle associated with this class is
given by
\begin{equation}
(p,q)=n(1,0)
\label{eq:77A}
\end{equation}
and, hence, $p=n$ and $q=0$.
It follows from~\eqref{eq:76} that the self--intersection number is
\begin{equation}                                                             
\vec{J} \cdot \vec{J}= -n^{2}
\label{eq:78}
\end{equation}
At this point, it is very useful to note from Figure $3$ that the projection
of such a membrane into the base produces a string indexed by $\vec{v}$,
integer $n$ and $(p,q)$. These are the fundamental objects that compose string
junction lattices \cite{bartonone,bartontwo}, which we review briefly in the
Appendix. Furthermore, the self--intersection number of class $\vec{v}$ given
in~\eqref{eq:76} coincides with the norm of the string $\vec{v}$ defined in
\cite{bartonone}. Therefore, computations from the points of view of the 
${\cal{T}}$--membrane and the string junction lattice coincide. We will,
therefore, use either method interchangeably. 
Note that, in general, membranes with boundary on the $(p,q)$ cycle
within a wrapped five--brane give rise to states with electric and magnetic
charges $p$ and $q$. As discussed in the Appendix, the condition for the 
associated state to be BPS saturated is that \cite{bartontwo,sethi}
\begin{equation}
\vec{J} \cdot \vec{J} \geq -2 + gcd(p,q)
\label{eq:79}
\end{equation}
where ``gcd'' stands for the greatest common positive 
divisor of $p$ and $q$. In the
case under consideration, since $q=0$, this condition becomes
\begin{equation}
\vec{J} \cdot \vec{J} \geq |n|-2
\label{eq:80}
\end{equation}
Comparing this with~\eqref{eq:78}, it is clear that the state will be BPS
saturated if and only if $n=\pm 1$. This CPT conjugate pair of 
allowed BPS states, $([1],[-1])$,
combine to form a single, stable  $N=2$ hypermultiplet $\Phi_{(1,0)}$, of 
electric charge $Q_{e}=1$ and vanishing magnetic charge, 
in the $M_{4}$ worldvolume theory of the wrapped five--brane. The non--BPS
states are either unstable or massive.

\onefigure{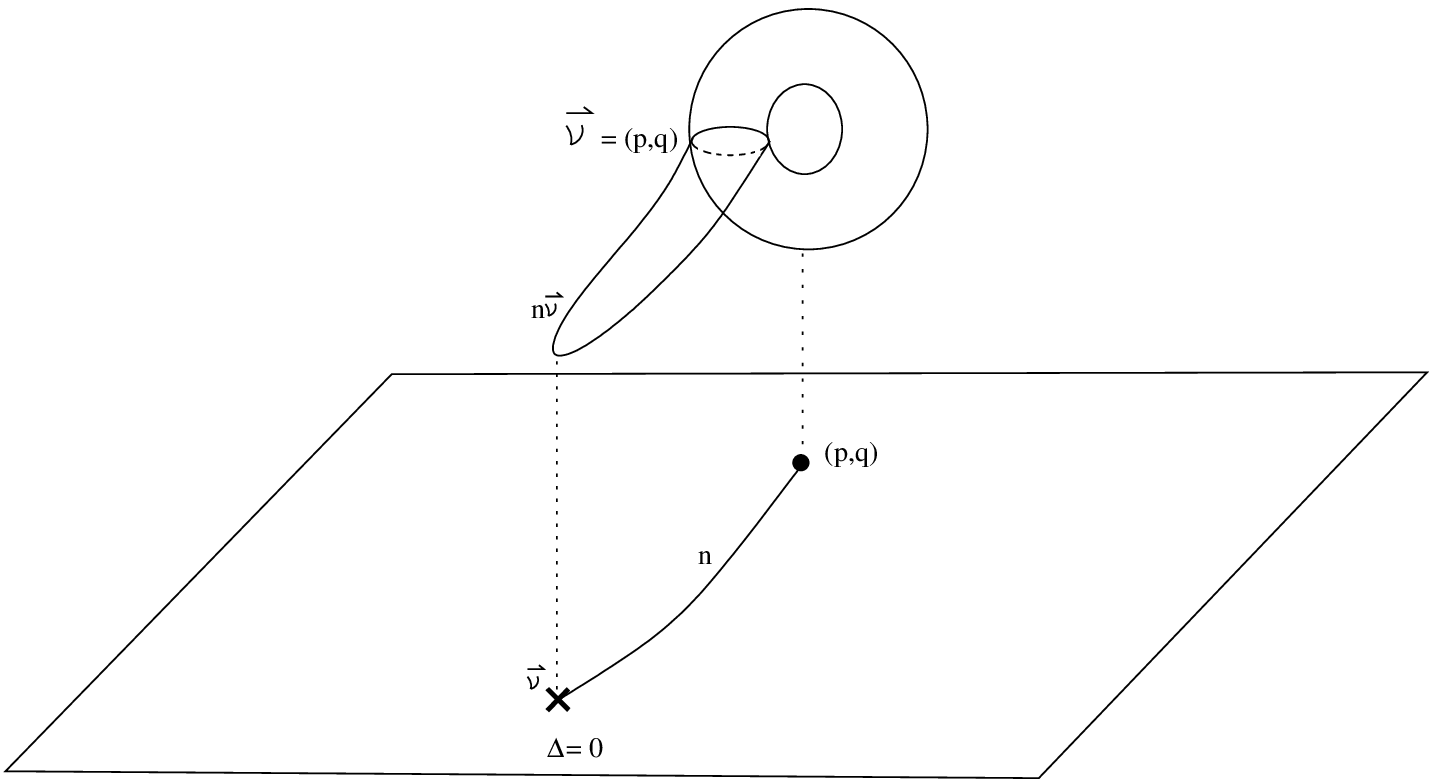}{A membrane with boundary cycle in
an elliptic fiber near a Kodaira type $I_1$ degeneracy.  In this example, the 
boundary cycle is the vanishing cycle $\vec{\cal{V}} =(p,q) = (n,0)$ 
associated with 
the $I_1$ fiber, and $n\vec{v}$ is the homology class $\vec{J}$ of the 
membrane.  
The discriminant locus is the point in the base at which $\Delta = 0$.
When projected to the base,  the membrane becomes the string illustrated
above.}

We conclude that, near a point on a smooth part of the discriminant curve with
elliptic fiber of Kodaira type $I_{1}$, the $M_{4}$ worldvolume theory of a
wrapped five--brane has $N=2$ supersymmetry at low energy. In addition
to the ``standard'' Abelian Yang--Mills supermultiplet with gauge connection 
$A_{\mu}$ and an uncharged hypermultiplet,
the $I_{1}$ degeneracy of the elliptic fiber produces a light BPS
hypermultiplet

\begin{itemize} 
\item $\Phi_{(1,0)}$   \qquad $Q_{e}=1$, $Q_{m}=0$
\end{itemize} 
This electric charge couples to the gauge connection $A_{\mu}$.  

The next simplest possibility, which, for example,
occurs at the smooth parts of the $\sigma$ component curve,  
is associated with Kodaira type $I_{2}$. We will discuss this case
in the next subsection.

\subsection*{Kodaira Type $I_{2}$:}

Pick any point on the smooth part of the discriminant curve with
a fiber of Kodaira type $I_{2}$, and choose the origin of the $u,v$
coordinates to be at that point. In the
neighborhood of this point, the associated sections have the form
\begin{equation}
g_{2}=u^{0}{\cal{G}}_{2}, \qquad g_{3}=u^{0}{\cal{G}}_{3}, \qquad
\Delta= u^{2}f^{2}
\label{eq:81}
\end{equation}
where ${\cal{G}}_{2}$, ${\cal{G}}_{3}$, and $f$, to leading order in
$u$, are non--zero functions of $v$ only (see \eqref{eq:54}). We see from~\eqref{eq:31A} that the
first two functions must satisfy~\eqref{eq:72}. It follows from~\eqref{eq:31}
and~\eqref{eq:81} that, at the origin, the Weierstrass form for the elliptic
fiber is again given by~\eqref{eq:73}. This is the standard Weierstrass
representation of an elliptic fiber of Kodaira type $I_{2}$. 
As discussed above, near such a
point, the $N=4$ supersymmetry of the low energy $M_{4}$ worldvolume theory of
the wrapped five--brane is reduced to $N=2$ supersymmetry. 
This is broken to $N=1$ supersymmetry by higher dimension operators.

As above, we can restrict the discussion to the elliptic fibration ${\cal{T}}$
over the path $\cP$ with coordinate $u$ that is transverse to
the discriminant curve . The discriminant curve now appears
as a point in the one--dimensional complex base $\cP$. Furthermore, the forms
of $g_{2}$, $g_{3}$ and $\Delta$, as well as the Weierstrass
form of the elliptic fiber, restricted to this twofold remains those given
in~\eqref{eq:81} and~\eqref{eq:73} with $\cG_{2}$, $\cG_{3}$ and $f$ evaluated at
$v=0$. Then it follows from Table $1$ that, within ${\cal{T}}$, the degeneracy
of the fiber near the discriminant point remains Kodaira type $I_{2}$.

The $SL(2, {\mathbb Z})$ monodromy transformation for a type $I_{2}$ fiber can be 
found from Table $1$, and is given by
\begin{equation}
{\cal{M}}_{I_{2}}=\begin{pmatrix}\
                  1 & 2 \\
                  0 & 1                                           
                  \end{pmatrix}.  
\label{eq:84}
\end{equation}
This transformation acts on the elements of $H_{1}(\cC_{2}, {\mathbb Z})$ and
has, up to multiplication by a non--zero integer, the single
eigenvector~\eqref{eq:75}
with eigenvalue $+1$. At this point, however, the situation diverges
substantially from that of the Kodaira type $I_{1}$ case above. This is
signaled by the appearance of the 2 in the monodromy matrix, which implies
that it can be decomposed as
\begin{equation}
{\cal{M}}_{I_{2}}= \begin{pmatrix}
                  1 & 1 \\
                  0 & 1
                  \end{pmatrix}
\cdot           \begin{pmatrix}
                  1 & 1 \\
                  0 & 1
                  \end{pmatrix}
                 =A \cdot A
\label{eq:86}
\end{equation}
It follows that there are now two copies of the eigenvector
$\vec{{\cal{V}}}=(1,0)$ that are relevant to the problem. 
One way to explore this
issue is to locally deform the Weierstrass equation and, hence. the discriminant 
in such a way that its locus, the single
point at the origin with Kodaira type $I_{2}$, is split into two 
nearby discriminant loci, which we label $1$ and $2$, 
each with Kodaira type $I_{1}$. Then, the monodromy at each of these points is
simply given by~\eqref{eq:74}, with a single eigenvector~\eqref{eq:75}. A
deformation of the Weierstrass equation that will accomplish this is given by
\begin{equation}
y^{2}=4x^{3} -{\cal{G}}_{2}x-{\cal{G}}_{3} +u(u+\epsilon)b
\label{eq:86A}
\end{equation}
Here, $b$ is the coefficient of the quadratic term in the $u$ expansion 
of $g_{3}$ and $\epsilon$ is a complex deformation parameter.  
Note, in passing, that under such a deformation a
$K3$ twofold remains $K3$. The meaning of this result is 
the following. 
Consider an elliptic fiber $\cC_{2}$ over a
point, $z$, near, but not at, the two discriminant loci. Then the one--cycle
$\vec{{\cal{V}}}=(1,0)$ is the unique cycle in $H_{1}(\cC_{2}, \mathbb Z)$
that contracts to zero as the fiber is moved to either of the two 
discriminant points. That
is, $\vec{{\cal{V}}}$ is the unique vanishing cycle associated with both
discriminant loci.

The physical implications arise from considering a ${\cal{T}}$--membrane 
whose boundary in the elliptic fiber over $z$ is $\vec{{\cal{V}}}=(1,0)$ and
which ``ends'' on the $I_{1}$ fiber over either
point $1$ or $2$ in the discriminant locus. Membranes 
of this type have a very specific
structure. Recall that in the $I_{1}$ case
above there was only one type of membrane, which ended
on the unique discriminant point. In contrast, in the $I_{2}$ case there 
are two types of membranes, one type ending on the $I_{1}$ fiber over 
discriminant point $1$ and the other type ending on the $I_{1}$ fiber over 
discriminant point $2$. We will denote the classes 
associated with discriminant points $1$ and $2$ by $\vec{v_{1}}$ and $\vec{v_{2}}$
respectively. Independently of whether or not ${\cal{T}}$ is a $K3$ surface,
we can show that the intersections of these membrane classes are given by
\begin{equation}
\vec{v_{1}} \cdot \vec{v_{1}}= \vec{v_{2}} \cdot \vec{v_{2}}=-1, \qquad
\vec{v_{1}} \cdot \vec{v_{2}}=0
\label{eq:87}
\end{equation}    
The proof for the self--intersections was presented in the section on $I_{1}$
and does not change here. Using the fact that $\vec{v}_{1}$ and $\vec{v}_{2}$
intersect only on the $(1,0)$ cycle in the fiber $\cC_{2}$ over $z$, and that
the self--intersection of this cycle vanishes, it follows that $\vec{v}_{1}
\cdot \vec{v}_{2}=0$.
A generic membrane of this type, projected into the base, is the string
junction shown in Figure $4$. We note that the membrane intersection numbers 
in~\eqref{eq:87} are identical to the string junction lattice
intersection form presented in~\eqref{eq:AA} and~\eqref{eq:A2} in the Appendix.
Therefore, once again, we can use the membrane and string junction lattice
formalism interchangeably. As the elliptic fiber approaches either of 
the discriminant points, the membrane
fluctuations become less and less heavy, becoming exactly massless when
the five--brane is wrapped on the degenerate $I_{1}$ Kodaira fiber over point
$1$ or $2$. Therefore, at either of the two discriminant loci, we expect
light BPS states to enter the $M_{4}$ low energy theory of the wrapped
five--brane.

To compute these states, we first note that any cycle of the form $(p,0)$,
where $p \in {\mathbb Z}$, is an eigenvector of the monodromy
${\cal{M}}_{I_{2}}$, not just $(1,0)$. The class of the associated membrane is
denoted by
\begin{equation}
\vec{J}= n_{1}\vec{v_{1}}+ n_{2}\vec{v_{2}}
\label{eq:88}
\end{equation}
where $n_{1},n_{2} \in {\mathbb Z}$. The boundary cycle associated with this
class is given by
\begin{equation}
(p,q)=n_{1}(1,0)+n_{2}(1,0)
\label{eq:88A}
\end{equation}
and, hence, $p=n_{1}+n_{2}$ and $q=0$.
It follows from~\eqref{eq:87} that the self--intersection number is
\begin{equation}
\vec{J} \cdot \vec{J}= -n_{1}^{2} -n_{2}^{2}
\label{eq:89}
\end{equation}
However, recall that~\eqref{eq:79} is the condition for the associated state
to be BPS saturated. 
In the case under consideration, since $q=0$, this condition becomes
\begin{equation}
\vec{J} \cdot \vec{J} \geq |n_{1}+n_{2}|-2
\label{eq:91}
\end{equation}
Comparing this with~\eqref{eq:89}, it is clear that a state will be BPS
saturated if and only if the pair $n_{1}, n_{2}$ takes the values $\pm 1,
0$ or 
$0, \pm 1$. The two BPS states described by $([1,0],[-1,0])$, combine to
form a single, stable $N=2$ hypermultiplet, $\Phi^{1}_{(1,0)}$, of electric
charge 
$Q_{e}=1$ and vanishing magnetic charge, 
in the $M_{4}$ worldvolume theory of the wrapped five--brane. 
Similarly, the
states associated with $([0,1],[0,-1])$ form a second, stable $N=2$
hypermultiplet,
$\Phi^{2}_{(1,0)}$, also with $Q_{e}=1$ and $Q_{m}=0$. 
The non-BPS states are either unstable or massive.

\onefigure{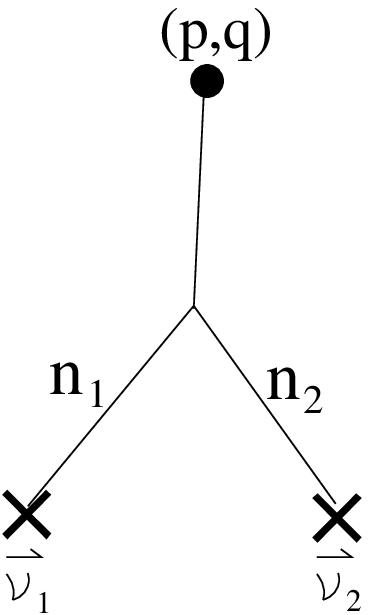}{A string junction arising for a split $I_2$ fiber.
The sum $n_1\vec{v}_1 + n_2\vec{v}_2$ is the homology class $\vec{J}$ 
of the membrane.
The boundary cycle is $(p,q)$, where $p = n_1 + n_2$ and $q = 0$.
The vanishing cycle associated with both $\vec{v}_1$ and $\vec{v}_2$ 
is $(1,0)$.} 

Further important information can be extracted by rewriting the membrane class
$\vec{J}$ in~\eqref{eq:88} as follows. Define classes 
\begin{equation}
\vec{w}_{p}=\frac{1}{2}(v_{1}+v_{2}), \qquad \vec{w}=\frac{1}{2}(v_{1}-v_{2})
\label{eq:92}
\end{equation}
Then, in terms of these classes, $\vec{J}$ can be written as 
\begin{equation}
\vec{J}=p\vec{w}_{p}+a\vec{w}
\label{eq:93}
\end{equation}
where $p=n_{1}+n_{2}$ and $a=n_{1}-n_{2}$. It is useful to proceed one step
further and write this equation as
\begin{equation}
\vec{J}=p\vec{w}_{p}+aC^{-1}\vec{\alpha}
\label{eq:94}
\end{equation}
where $C^{-1}=\frac{1}{2}$ and $\vec{\alpha}=2\vec{w}$. Note that 
\begin{equation}
\vec{\alpha}=v_{1}-v_{2}
\label{eq:95}
\end{equation}
which is associated with the cycle $(p,q)=(0,0)$ and, hence,
corresponds to no boundary cycle at all. Such classes are described by curves from
discriminant point $1$ to discriminant point $2$, as shown in Figure $5$.
Furthermore, it follows from~\eqref{eq:87} that 
\begin{equation}
\vec{\alpha} \cdot \vec{\alpha}=-2
\label{eq:96}
\end{equation}
We see, by comparing equations~\eqref{eq:94} and~\eqref{eq:95} with 
\eqref{eq:A4} and \eqref{eq:A5}
in the Appendix, that class $\vec{\alpha}$ corresponds to the simple 
root of the Lie
algebra of $SU(2)$. 

\onefigure{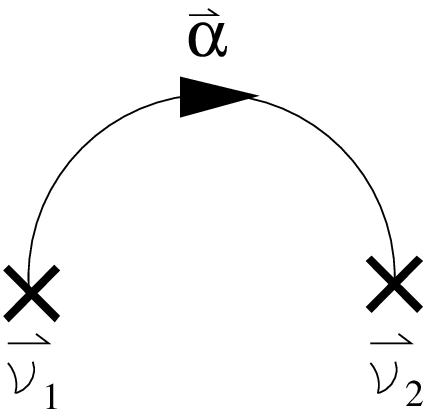}{The junction corresponding to the simple root of $SU(2)$.}

Note that, written as a pair $[n_{1}, n_{2}]$, class
$\vec{\alpha}= [1,-1]$. 
It follows that the BPS states that make up the hypermultiplet
$\Phi^{1}_{(1,0)}$,
specified by $[1,0]$ and $[-1,0]$, are related to the BPS states, 
$[0,1]$ and $[0,-1]$, that make up the hypermultiplet $\Phi^{2}_{(1,0)}$
through addition and subtraction of the root vector $\vec{\alpha}$. That is
\begin{equation}
[1,0] -\vec{\alpha} =[0,1], \qquad [-1,0] +\vec{\alpha} =[0,-1]
\label{eq:96A}
\end{equation}
Further addition and subtraction of the root $\vec{\alpha}$ leads to
unstable or massive non--BPS states. It follows that hypermultiplets
$\Phi^{1}_{(1,0)}$ and $\Phi^{2}_{(1,0)}$ form a doublet representation 
of $SU(2)$. The
appearance of an $SU(2)$ global group can be read off directly 
from the A-D-E column of Table $1$. For
Kodaira type $I_{2}$, the A-D-E classification is $A_{1}$, which corresponds
to the group $SU(2)$.

We conclude that, near a point on a smooth part of the discriminant curve with
elliptic fiber of Kodaira type $I_{2}$, the $M_{4}$ worldvolume theory of a
wrapped five--brane has $N=2$ supersymmetry at low energy. In addition to the
``standard'' Abelian Yang--Mills supermultiplet with gauge connection
$A_{\mu}$ and an uncharged hypermultiplet, the $I_{2}$ degeneracy of the
elliptic fiber produces an $SU(2)$ doublet ${\bf 2 \rm}$ 
of light BPS hypermultiplets

\begin{itemize}
\item  $\Phi^{i}_{(1,0)}$   \qquad $Q_{e}=1, Q_{m}=0$
\end{itemize}
where $i=1,2$. The electric charge couples to the gauge connection $A_{\mu}$.

We now turn our attention to more a more complicated situation, specifically
that of a Kodaira type $IV$ fiber occuring at the smooth part of the
discriminant curve. Such fibers are found, for example, 
over the smooth part of the 
component curve $\cS$ in Case $3$ above. The analysis of this example
contains most 
of the elements necessary to compute the light BPS states for a fiber of
any Kodaira type over a smooth part of the discriminant curve.

\subsection*{Kodaira Type IV:}

Pick any point on the smooth part of the discriminant curve with a
fiber of Kodaira type $IV$, and choose the origin of the $u,v$ coordinates
to be at that point. In the neighborhood of this point, the associated 
sections have the form
\begin{equation}
g_{2}=u^{L}{\cal{G}}_{2}, \qquad g_{3}=u^{2}{\cal{G}}_{3}, \qquad
\Delta= u^{4}f^{2}
\label{eq:98}
\end{equation}
where $L>1$ and ${\cal{G}}_{2}$, ${\cal{G}}_{3}$ and 
$f$, to leading order in $u$, are non--vanishing functions of $v$ only. 
Further analysis of the explicit curve $\cS$ in Case $3$ above fixes the value
of coefficient $L$ to be $L=2$. 
We see from~\eqref{eq:31A} that, near the discriminant locus,
\begin{equation}
\Delta= g_{2}^{3}-27g_{3}^{2}=u^{4}(-27{\cal{G}}_{3}^{2})
\label{eq:99}
\end{equation}
which vanishes at the locus without any further restrictions on
${\cal{G}}_{2}$ and ${\cal{G}}_{3}$. It follows from~\eqref{eq:31}
and~\eqref{eq:98} that, at the origin, the Weierstrass form of the elliptic
fiber becomes
\begin{equation}
y^{2}=4x^{3}-u^{2}{\cal{G}}_{2}x-u^{2}{\cal{G}}_{3}
\label{eq:100}
\end{equation}
where we have used affine coordinates with $z=1$. This is the 
Weierstrass representation of an elliptic fiber of Kodaira type $IV$.
Near such a point, the $N=4$ supersymmetry of the low energy $M_{4}$ 
worldvolume theory of the wrapped five--brane is reduced to $N=2$ supersymmetry.  
This is broken to $N=1$ supersymmetry by higher dimensional operators.

As above, we can restrict the discussion to the elliptic fibration 
over the path $\cP$ with coordinate $u$ that is transverse to
the discriminant curve . The discriminant curve now appears
as a point in the one--dimensional complex base $\cP$. The forms of
$g_{2}$, $g_{3}$ and $\Delta$, as well as the Weierstrass
form of the elliptic fiber, restricted to this twofold remain those given
in~\eqref{eq:98} and~\eqref{eq:100} with $\cG_{2}$, $\cG_{3}$ and $f$
evaluated at $v=0$. It follows from Table $1$ that, within ${\cal{T}}$, the
degeneracy of the fiber near the discriminant point remains that of Kodaira
type $IV$.

The $SL(2, {\mathbb Z})$ monodromy 
transformation for a type $IV$ fiber can be found from Table $1$. Here, for
technical reasons, it is convenient to use a  monodromy matrix conjugate to
the one listed in Table $1$ given by
\begin{equation}
{\cal{M}}_{IV}= \begin{pmatrix}
                  -1 & 1 \\
                  -1 & 0
                  \end{pmatrix}
\label{eq:101}
\end{equation}
This transformation acts on the elements of $H_{1}(\cC_{2}, {\mathbb Z})$.
However, unlike the previous cases of Kodaira type $I_{1}$ and $I_{2}$, 
${\cal{M}}_{IV}$ has no real eigenvector and, therefore, 
there is no obvious associated vanishing cycle. An indication as how to
proceed is given by the fact that ${\cal{M}}_{IV}$ can be decomposed as
\begin{equation}
{\cal{M}}_{IV}=  \begin{pmatrix}
                  1 & 1 \\
                  0 & 1
                  \end{pmatrix}
\cdot  
 \begin{pmatrix}
                  1 & 1 \\
                  0 & 1
                  \end{pmatrix}
\cdot 
 \begin{pmatrix}
                  1 & 1 \\
                  0 & 1
                  \end{pmatrix}
\cdot
 \begin{pmatrix}
                  2 & 1 \\
                  -1 & 0
                  \end{pmatrix}
 =A\cdot A\cdot A\cdot B
\label{eq:102}
\end{equation}
Following the technique employed in the $I_{2}$ case, we proceed by deforming the
discriminant curve in such a way that its locus, the single point at the
origin with Kodaira type $IV$, is split into four nearby discriminant points,
which we label $1$,$2$,$3$ and $4$. This corresponds to a relevant 
deformation of the Weierstrass representation given by
\begin{equation}
y^{2}=4x^{3}- u^{2}{\cal{G}}_{2}x- u^{2}{\cal{G}}_{3}+\epsilon
\label{eq:102A}
\end{equation}
where $\epsilon$ is a constant deformation parameter.  
The first three points each have a  
Kodaira fiber of type $I_{1}$ with the monodromy given in~\eqref{eq:74}
and, up to multiplication by a non--zero integer, the single
eigenvector~\eqref{eq:75}.
The fourth point, however, has monodromy
\begin{equation}
B= \begin{pmatrix}
   2 & 1  \\
  -1 & 0
   \end{pmatrix}
\label{eq:105}
\end{equation}
with, up to multiplication by a non--zero integer, the eigenvector
\begin{equation}
\vec{\tilde{\cal{V}}}=(1,-1)
\label{eq:106}
\end{equation}
At first glance, it is not obvious what Kodaira type one is finding in this
last case. However, it is easy to show that $A$ is,
in fact,
conjugate to $B$ and, hence, the fiber at point $4$ is also of
Kodaira type $I_{1}$. The meaning of these results is the following.
Consider an elliptic fiber $\cC_{2}$ over a point, $z$, near, but not at, the four
discriminant loci. Then the one--cycle $\vec{{\cal{V}}}=(1,0)$ 
is the unique cycle in $H_{1}(\cC_{2}, \mathbb Z)$ that
contracts to zero as the fiber is moved to each of the
discriminant points $1$, $2$, and $3$ without encircling $4$. That
is, $\vec{{\cal{V}}}$ is the unique vanishing cycle associated with the first
three discriminant loci. Similarly, the one--cycle
$\vec{\tilde{\cal{V}}}=(1,-1)$ 
is the unique cycle in $H_{1}(\cC_{2}, \mathbb Z)$ that
contracts to zero as the fiber is moved to the discriminant point $4$ 
without encircling $1, \ 2$ or $3$.
That is, $\vec{\tilde{\cal{V}}}$ is the unique vanishing cycle associated with
the fourth discriminant locus.

The physical implications of this arise as follows. ${\cal{T}}$-- membranes that
are bounded by the vanishing cycle $\vec{{\cal{V}}}=(1,0)$ in the elliptic
fiber over $z$ are of a very specific
structure. There 
are three membranes of this type, each ``ending'' on the $I_{1}$ fiber over 
discriminant points $1$,$2$ or $3$ respectively. Similarly, membranes that
are bounded by the vanishing cycle $\vec{\tilde{\cal{V}}}=(1,-1)$ have a very
specific
structure. These membranes ``end'' on the $I_{1}$ fiber over the 
discriminant point $4$.  
We will denote the ${\cal{T}}$--membrane classes 
associated with
discriminant points $1$, $2$,$3$ and $4$ by $\vec{v_{1}}$,$\vec{v_{2}}$,
$\vec{v_{3}}$ and $\vec{v_{4}}$
respectively. A generic membrane class of this type, 
projected into the base, is the string
junction shown in Figure $6$.
Note from the Appendix that the intersections of these membrane
classes are given by  
\begin{equation}
\vec{v_{1}} \cdot \vec{v_{1}}= \vec{v_{2}} \cdot \vec{v_{2}}=
\vec{v_{3}} \cdot \vec{v_{3}}=\vec{v_{4}} \cdot \vec{v_{4}}=-1
\end{equation}
\begin{equation}
\vec{v_{1}} \cdot \vec{v_{2}}=\vec{v_{1}} \cdot \vec{v_{3}}= 
\vec{v_{2}} \cdot \vec{v_{3}}= 0
\label{eq:107}
\end{equation}
\begin{equation}
\vec{v_{1}} \cdot \vec{v_{4}}=\vec{v_{2}} \cdot \vec{v_{4}}= 
\vec{v_{3}} \cdot \vec{v_{4}}= -\frac{1}{2}
\end{equation}
As the elliptic fiber approaches any one of 
the discriminant points, the membrane
fluctuations become less and less heavy, becoming exactly massless when
the five--brane is wrapped on the degenerate $I_{1}$ Kodaira fiber over point
$1$,$2$,$3$ or $4$. Therefore, at any of the four discriminant loci, we expect
light BPS states to enter the $M_{4}$ low energy theory of the wrapped
five--brane.

To compute these states, we first note, by analogy with the above
discussion, that 
the classes of the associated membranes are given by
\begin{equation}
\vec{J}= \Sigma^{4}_{i=1}n_{i}\vec{v_{i}}
\label{eq:108}
\end{equation}
where $n_{i}$, $i=1,..,4$ are arbitrary integers. Using~\eqref{eq:75}
and~\eqref{eq:106}, it follows that the associated boundary cycles in
$H_{1}(\cC_{2}, \mathbb Z)$ are given by
\begin{equation}
(p,q)= \Sigma_{i=1}^{3}n_{i}(1,0) +n_{4}(1,-1)
\label{eq:108A}
\end{equation}
and, hence, $p=\Sigma_{i=1}^{4}n_{i}$ and $q=-n_{4}$.
From~\eqref{eq:107}, we find that the self--intersection number of class
$\vec{J}$ is
\begin{equation}
\vec{J} \cdot \vec{J}= -\Sigma_{i=1}^{4}n_{i}^{2}- (n_{1}+n_{2}+n_{3})n_{4}  
\label{eq:109}
\end{equation}
However, recall that~\eqref{eq:79} is the condition for the associated state
to be BPS saturated. Comparing this to~\eqref{eq:109}, it was shown in
\cite{bartonone,bartontwo} that a state will
be BPS saturated if and only if $n_{1},n_{2},n_{3},n_{4}$ take the 
values

\vspace{30pt}

\begin{tabular}{|l|c|c} 
\hline
$(n_{1}, n_{2}, n_{3}, n_{4})$  & $(p,q)$\\
\hline
$(1,0,0,0), (0,1,0,0), (0,0,1,0)$ & $(1,0)$\\

$(-1,0,0,0), (0,-1,0,0), (0,0,-1,0)$ & $(-1,0)$\\
\hline
$(1,0,0,-1), (0,1,0,-1), (0,0,1,-1)$ & $(0,1)$\\

$(-1,0,0,1), (0,-1,0,1), (0,0,-1,1)$ & $(0,-1)$\\
\hline
$(0,-1,-1,1), (-1,0,-1,1), (-1,-1,0,1)$ & $(-1,-1)$\\

$(0,1,1,-1), (1,0,1,-1), (1,1,0,-1)$ & $(1,1)$\\
\hline
$(1,1,1,-2)$ & $(1,2)$\\

$(-1,-1,-1,2)$ & $(-1,-2)$\\
\hline
$(0,0,0,1)$ & $(1,-1)$\\

$(0,0,0,-1)$ & $(-1,1)$\\
\hline
$(1,1,1,-1)$ & $(2,1)$\\

$(-1,-1,-1,1)$ & $(-2,-1)$\\
\hline
\end{tabular}

\vspace{30pt}

Table 2: The BPS states associated with a fiber of Kodaira type $IV$.

\vspace{30pt}

The six BPS states in the first row combine to form a triplet of $N=2$
hypermultiplets, each of electric charge $Q_{e}=1$ and vanishing 
magnetic charge, in the $M_{4}$ worldvolume
theory of the wrapped five--brane. We denote these hypermultiplets by
$\Phi_{(1,0)}^{i}$, where $i=1,2,3$. Similarly, the second and third rows each
correspond to a triplet of $N=2$ hypermultiplets, which we denote by
$\Phi_{(0,1)}^{i}$ and $\Phi_{(-1,-1)}^{i}$ with $i=1,2,3$ 
respectively. As indicated
by the notation, the multiplet $\Phi_{(0,1)}^{i}$ has vanishing electric charge
and magnetic charge $Q_{m}=1$, whereas $\Phi_{(-1,-1)}^{i}$ carries both electric
charge $Q_{e}=-1$ and magnetic charge $Q_{m}=-1$. The two BPS
states in the fourth row combine to form a single,
$N=2$ hypermultiplet, $\Psi_{(1,2)}$ with $Q_{e}=1$ and $Q_{m}=2$. 
Finally, the last two rows each correspond to a 
single $N=2$ hypermultiplet, denoted by 
$\Psi_{(1,-1)}$ and $\Psi_{(2,1)}$, with $Q_{e}=1$, $Q_{m}=-1$ and 
$Q_{e}=2$, $Q_{m}=1$
respectively. The non--BPS states are either unstable or massive.
When the deformation that split the $IV$ fiber is undone, all these BPS 
multiplets
become simultaneously massless as the fivebrane is moved towards the 
$IV$ fiber.
These states have mutually non-local $(p,q)$ charges,  
meaning that they can not 
be made simultaneously purely electric by an $SL(2, {\mathbb Z})$ 
transformation. 
This leads to a very exotic low energy theory without a local Lagrangian 
description. Such theories were 
obtained originally, in a different context, in 
\cite{argyresdouglas, argwitten, minahan}.
When the fivebrane wraps the type $IV$ Kodaira fiber,  the low energy theory
flows to an interacting fixed point in the infrared.

\onefigure{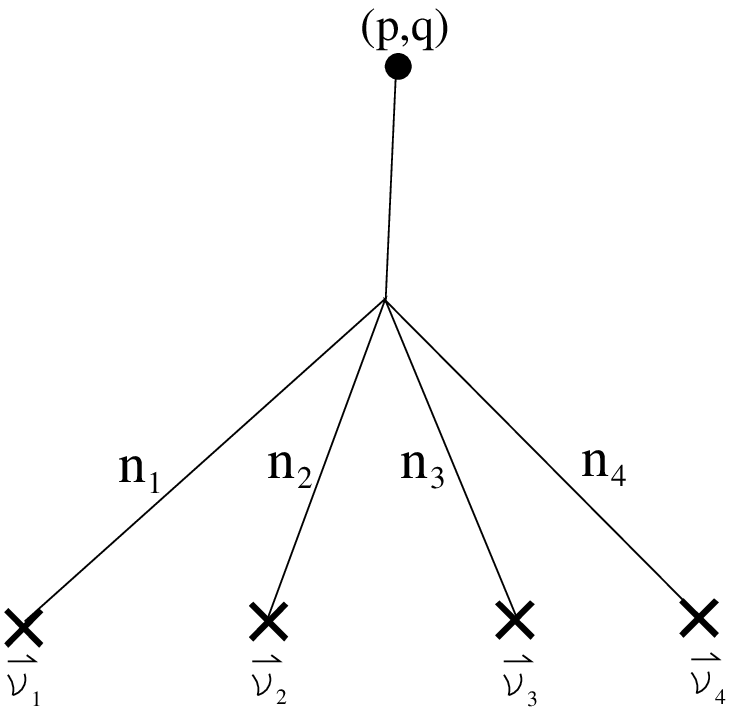}{A string junction for a split type IV Kodaira fiber.
The sum $\sum_{i=1}^4 n_i\vec{v}_i$ is the homology class $\vec{J}$ of the 
membrane, and $(p,q)$
is the boundary cycle. In this case $p = \sum_{i=1}^4 n_i$ and $q = -n_4$
The vanishing cycle associated with $\vec{v}_1, \vec{v}_2$ and $\vec{v}_3$
is $(1,0)$,  while that of the $\vec{v}_4$ is $(1,-1)$.}

Further important information can be extracted by rewriting the membrane class
$\vec{J}$ as follows. Define classes
\begin{equation}
\vec{w}_{p}= \frac{1}{3}(v_{1}+v_{2}+v_{3}) , \qquad  
\vec{w}_{q}=\frac{1}{3}(v_{1}+v_{2}+v_{3})-v_{4}
\label{eq:111}
\end{equation}
and
\begin{equation}
\vec{w}^{1}= \frac{1}{3}(2v_{1}-v_{2}-v_{3}) , \qquad  
\vec{w}^{2}= \frac{1}{3}(v_{1}+v_{2}-2v_{3})
\label{eq:112}
\end{equation}
Then, in terms of these classes, $\vec{J}$ can be written as 
\begin{equation}
\vec{J}= p\vec{w}_{p} +q\vec{w}_{q}+ a_{i}\vec{w}^{i}
\label{eq:113}
\end{equation}
where $p$ and $q$ are given in~\eqref{eq:108A} and 
\begin{equation}
a_{1}= n_{1}-n_{2},  \qquad a_{2}= n_{2}-n_{3}
\label{eq:114}
\end{equation}
It is useful to proceed one step further and write equation~\eqref{eq:113} as 
\begin{equation}
\vec{J}= p\vec{w}_{p} +q\vec{w}_{q}+ a_{i}C^{-1ij}\vec{\alpha}_{i}
\label{eq:115}
\end{equation}
where $\vec{\alpha}_{1}=2w^{1}-w^{2}$, $\vec{\alpha}_{2}=-w^{1}+2w^{2}$ and
$C_{ij}=-\vec{\alpha}_{i} \cdot \vec{\alpha}_{j}$.  Note that
\begin{equation}
\vec{\alpha}_{1}=v_{1}-v_{2}, \qquad 
\vec{\alpha}_{2}=v_{2}-v_{3}
\label{eq:116}
\end{equation}
which are both associated with the cycle $(p,q)=(0,0)$ and, hence, 
each corresponds to no boundary cycle at all. Therefore,
$\vec{\alpha}_{1}$ is described by curves from discriminant point $1$ to
discriminant point $2$ and $\vec{\alpha}_{2}$ is described by 
curves from discriminant point $2$ to discriminant point $3$, as shown
in Figure $6$. Furthermore, it follows from~\eqref{eq:107} that
\begin{equation}
C_{ij}= \begin{pmatrix}
   2 &-1  \\
  -1 & 2
   \end{pmatrix}
\label{eq:117}
\end{equation}
which is the Cartan matrix of the Lie algebra of $SU(3)$. Hence, the classes
$\vec{\alpha}_{1}$ and $\vec{\alpha}_{2}$ correspond to the simple roots of
the $SU(3)$. Note that, written as a four--tuplet $[n_{1},n_{2},n_{3},n_{4}]$,
class $\vec{\alpha}_{1}=[1,-1,0,0]$ and class
$\vec{\alpha}_{2}=[0,1,-1,0]$. 

\onefigure{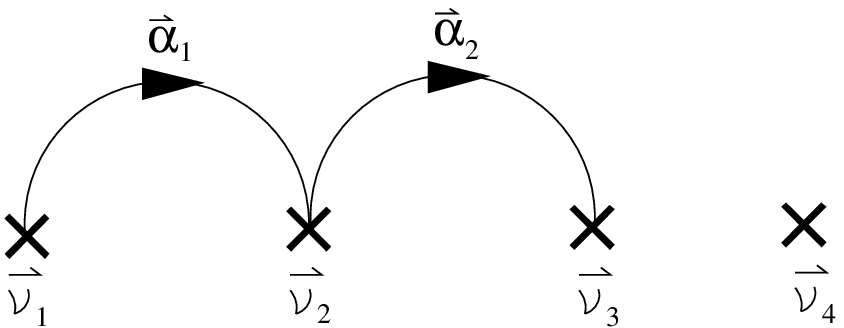}{The two simple roots of $SU(3)$, realized as 
string junctions 
for the split Kodaira type IV fiber.  Note that there are no root 
junctions with
an endpoint on the fourth $I_1$ fiber.}

Consider, for example, the six BPS states in
the first row of Table $2$, and note that
\begin{equation}
[1,0,0,0]-\vec{\alpha}_{1}= [0,1,0,0], \qquad 
[0,1,0,0]-\vec{\alpha}_{2}= [0,0,1,0]
\end{equation}
\begin{equation}
[-1,0,0,0]+\vec{\alpha}_{1}= [0,-1,0,0], \qquad 
[0,-1,0,0]+\vec{\alpha}_{2}= [0,0,-1,0]
\label{eq:118}
\end{equation}
Further addition or subtraction of the roots leads to non--BPS states that are
either unstable or massive. 
Hence, the pairs of BPS states $([1,0,0,0], [-1,0,0,0])$, 
$([0,1,0,0], [0,-1,0,0])$ and $([0,0,1,0], [0,0,-1,0])$ combine to form 
the hypermultiplets $\Phi_{(1,0)}^{1},\Phi_{(1,0)}^{2}$ and
$\Phi_{(1,0)}^{3}$ respectively. Furthermore, these hypermultiplets 
form a triplet $\bf 3 \rm$ representation of $SU(3)$. The same is
true for the hypermultiplets $\Phi_{(0,1)}^{i}$ and $\Phi_{(-1,-1)}^{i}$ for
$i=1,2,3$, each of which transforms as an $SU(3)$ $\bf 3 \rm$ representation.
Now consider the two BPS states in the fourth row of Table $2$. Addition or
subtraction of any root leads immediately to non--BPS states which are either
unstable of massive. Hence,
the pair of BPS states $([1,1,1,-2], [-1,-1,-1,2])$ combine to form a
hypermultiplet $\Psi_{(1,2)}$, which is a singlet under $SU(3)$. The same
is true
for $\Psi_{(1,-1)}$ and $\Psi_{(2,1)}$ which are both $SU(3)$ singlets.
The appearance of the global group $SU(3)$ can be read off directly from the 
A-D-E column of Table $1$. For Kodaira type $IV$, the A-D-E classification 
is $A_{2}$, which corresponds to the group $SU(3)$.

We conclude that, near a point on a smooth part of the discriminant curve with
elliptic fiber of Kodaira type $IV$, the $M_{4}$ worldvolume theory of a
wrapped five--brane has $N=2$ supersymmetry at low energy. In addition to the
``standard'' Abelian Yang--Mills supermultiplet with gauge connection
$A_{\mu}$ and an uncharged hypermultiplet, the $IV$ degeneracy of the
elliptic fiber produces light BPS hypermultiplets. These fall into $SU(3)$
$\bf 3 \rm$ representations
\begin{itemize}
\item $\Phi_{(1,0) }^{i}$   \qquad $Q_{e}=1, Q_{m}=0$
\end{itemize}
\begin{itemize}
\item $\Phi_{(0,1)}^{i}$   \qquad $Q_{e}=0, Q_{m}=1$
\end{itemize}
\begin{itemize}
\item $\Phi_{(-1,-1)}^{i}$   \qquad $Q_{e}=-1, Q_{m}=-1$
\end{itemize}
with $i=1,2,3$ and $SU(3)$ singlets
\begin{itemize}
\item $\Psi_{(1,2)}$   \qquad $Q_{e}=1, Q_{m}=0$
\end{itemize}
\begin{itemize}
\item $\Psi_{(1,-1)}$   \qquad $Q_{e}=1, Q_{m}=-1$
\end{itemize}
\begin{itemize}
\item $\Psi_{(2,1)}$   \qquad $Q_{e}=2, Q_{m}=1$
\end{itemize}
The electric charge couples to gauge connection $A_{\mu}$ whereas the magnetic
charge couples to $\tilde{A_{\mu}}$ defined by $d\tilde{A_{\mu}}=*F$.

We end this section by presenting the BPS states associated with
each of the remaining Kodaira types, $I_{0}^{*}$ ,$III^{*}$, and $IV^{*}$, 
over the smooth parts of the discriminant curve in the $B=\hat{\mathbb F}_{3}$
model.

\subsection*{Kodaira Types $I_{0}^{*}$, $III^{*}$, and $IV^{*}$:}

The A-D-E symmetry algebra associated with a fiber of Kodaira type $I_{0}^{*}$
can be read off from Table $1$ and is given by $D_{4}$.
The associated global symmetry group is $SO(8)$.
The low energy theory that arises in the neighborhood of an $I_{0}^{*}$ 
fiber is the same as that of an $N=2$, $SU(2)$ Yang--Mills theory with four 
quark flavors. The BPS
multiplets in the $I_{0}^{*}$ case can be easily found by comparing 
to Yang-Mills theory results \cite{ferrari1, ferrari2}, or by using string 
junctions as in \cite{bartontwo}. The results are summarized in the table below.

\vspace{30pt}

\begin{tabular}{|c|c|} 
\hline
$(p,q)$ charges  & $SO(8)$ representation\\
\hline
$(2n,2m)$ & ${\bf 1 \rm}$\\
\hline
$(2n+1,2m)$ &  ${\bf 8_v \rm}$\\
\hline
$(2n,2m+1)$ & ${\bf 8_s \rm}$\\
\hline
$(2n+1,2m+1)$ & ${\bf 8_c \rm}$\\
\hline
\end{tabular}

\vspace{30pt}

Table 3: The BPS multiplets associated with a fiber of Kodaira type
$I_{0}^{*}$.

\vspace{30pt}

\noindent Table $3$ is to be read as follows. For each $n$ and $m$, there is
an $N=2$ multiplet with the $(p,q)$ charges listed in the first column,
which transforms in the representation of $SO(8)$ listed in 
the last column.
In the first row,  $n$ and $m$ are constrained to be relatively prime,
whereas in the remaining rows, $p$ and $q$ must be relatively prime.
There are no further constraints. 
Note that, in analogy to the Kodaira type $IV$ case, there are a finite 
number of representations of $SO(8)$ which appear, specifically singlets and 
octets only. However, unlike the type $IV$ case, each of these
representations occurs with infinite multiplicity.  
Of course not all of these multiplets are simultaneously stable, depending on 
the location of the five--brane in moduli space. 

Let us summarize these results in terms of $N=2$ supermultiplets.
In addition to the ``standard'' Abelian vector supermultiplet with gauge
connection $A_{\mu}$ and an uncharged hypermultiplet, the $I_{0}^{*}$
degeneracy of the elliptic fiber produces light BPS hyper- and vector
multiplets. The extra vector multiplets are $SO(8)$ singlets with $p=\pm 2$
and $q=0$
\begin{itemize}
\item $V_{\pm}$
\end{itemize}
As the five--brane approaches the discriminant curve, these combine with the
usual uncharged Abelian vector multiplet to form massless $N=2$ vector
multiplets $V_{a}$ which transform as the adjoint representation of an
enhanced $SU(2)$ gauge group.
The remaining states belong to hypermultiplets
\begin{itemize}
\item $\Phi^{i}_{(p,q)}$ \qquad  $Q_{e}=p$,  $Q_{m}=q$
\end{itemize}
where the charges $(p,q)$ and the $SO(8)$ representation multiplets $i$ are
given in Table $3$. As the five--brane approaches the discriminant curve, the
hypermultiplets transforming as ${\bf 8_{v} \rm}$ under $SO(8)$ with $q=0$
combine to form an $SU(2)$ doublet ${\bf 2 \rm}$, $Q^{i}_{A}$, where $i$ is
the index of the ${\bf 8_{v} \rm}$ representation and $A=1,2$. These
correspond to hypermultiplets of four $SU(2)$ doublet quark flavors.

The A-D-E symmetry algebras associated with fibers of Kodaira type $III^{*}$
and $IV^{*}$ can be read off from Table $1$. The associated global symmetry
groups are the exceptional groups $E_{7}$ and $E_{6}$ respectively. 
The complete spectrum of BPS multiplets for these two Kodaira types are much
harder to determine, for reasons discussed below, and we will present only
partial results \cite{bartontwo}. 
First, note from Table $4$ in the Appendix that the monodromy
associated with a Kodaira fiber of type $I_{0}^{*}$ decomposes into type
$I_{1}$ monodromies as $AAAABC$. Similarly, one sees that Kodaira fibers of
type $III^{*}$ and $IV^{*}$ decompose as $AAAAAABCC$ and $AAAAABCC$
respectively. It follows that the $I_0^*$ string junction lattice is a
sublattice 
of the $III^{*}$ and $IV^{*}$ junction lattices.
This reflects the fact that a fiber of type $I_0^*$ may be obtained from 
type $III^*$ or type $IV^*$ fibers by deformations of the discriminant curve; 
one simply moves some of the $I_1$ fibers in the decomposition of the
$III^*$ or
$IV^*$ to infinity, leaving only the $I_1$ fibers which make up the
$I_0^*$.  For instance, the $IV^*$ fiber decomposes into $I_1$ fibers
of monodromy type AAAAABCC.  By moving one $I_1$ of type A
and another of type $C$ to infinity,  one is left with $I_{1}$ fibers of
monodromy type AAAABC,  which is the decomposition of an $I_0^*$ fiber.
It follows from this that each of the $I_0^*$ BPS multiplets listed in 
Table $3$ is also a BPS multiplet of both
the Kodaira type $III^*$ and type $IV^*$ fibers. However,
it turns out that $I_0^*$ multiplets with different $(p,q)$ charges
transforming
in the same representation of $SO(8)$, transform as different 
representations of the exceptional groups of the Kodaira fibers $III^*$ and
$IV^*$. For example \cite{bartontwo}, 
the ${\bf 8_v \rm}$ appearing for the $I_0^*$ fiber
with charges $(p,q) = (1,0)$ is embedded in a ${\bf 56 \rm}$ of $E_7$, or a 
${\bf 27 \rm}$ of $E_6$, 
while the ${\bf 8_v \rm}$ with charges $(p,q) = (1,2)$ is embedded in a 
${\bf 27664 \rm}$ of $E_7$, or a ${\bf 351 \rm}$ of $E_6$.
Consequently, although the type $I_0^*$ BPS multiplets with $(p,q)$ charges
listed in Table $3$ are also BPS multiplets of type $III^*$ and type $IV^*$
fibers, they are classified by an infinite number of different $E_7$ and $E_6$
representations. Although these can be computed on a case by case basis, a
simple listing of such multiplets is impossible.
In addition, there are BPS multiplets associated with both type 
$III^*$ and type $IV^*$ fibers that are unrelated to those of the 
$I_0^*$. These multiplets
arise from string junctions involving the type $A$ and type $C$ $I_{1}$ fibers
not contained in the decomposition of $I_0^*$. We will not discuss these
states here, referring the interested reader to \cite{bartontwo}.
Again, one expects BPS states with mutually non-local charges to become
simultaneously massless as the five-brane approaches the singular fiber.
The low energy theory on the five-brane wrapping the singular fiber
flows to an exotic interacting fixed point theory.  The fixed point
theories with exceptional global symmetries were first discussed,
in a different context,  in \cite{minahan}.


\section{Conclusion:}


In this paper, we have presented detailed techniques for computing the
discriminant curves of elliptically fibered Calabi--Yau threefolds. These were
applied to a specific three--family, $SU(5)$ GUT model of particle physics 
within the context of
Heterotic M--Theory. In this theory, and in general, the discriminant curves
have an intricate structure, consisting of smooth sections, cusps and
tangential and normal self--intersections. The type of degeneration of the
elliptic fiber, classified by Kodaira, changes in the different 
regions of the discriminant curve. In this paper, we discussed how to find the Kodaira
type of the fiber singularities and explicitly computed them for the
discriminant curves associated with the $SU(5)$ GUT model. In Heterotic M--Theory,
anomaly cancellation generically requires the existence of five--branes,
located in the bulk space, wrapped on a holomorphic curve in the associated
Calabi--Yau threefold. We showed that there is always a region of the moduli
space of this holomorphic curve that corresponds a single five--brane wrapped
on a pure fiber elliptic curve. Since this fiber degenerates as it approaches
the discriminant curve, one expects light BPS states to appear in the
worldvolume theory of this five--brane. For points on the smooth parts of the
discriminant curve, we demonstrated in detail how the M--theory membranes
associated with these states are, when projected into the base space,
related to string junction lattices. Using string junction techniques, we
computed the massless BPS 
hyper- and vector multiplets for all the Kodaira type
degeneracies that occur in our specific $SU(5)$ GUT model. The Kodaira theory,
as well as the computation of light states, is considerably more intricate at
the cusp and self--intersection points of the discriminant curve. These topics
will be discussed in a forthcoming publication.

It is important to note that the solutions of the BPS constraint on string 
junctions only gives a list of the possible BPS states in the spectrum.  
Determining their stability at different points in moduli space is a dynamical
question,  and the answer is not completely known for the exceptional Kodaira types. 
The answer is known for the non--exceptional Kodaira degeneracies,
including the types $I_{1}$, $I_{2}$, $IV$ and $I_{0}^{*}$ discussed in this paper 
\cite{ferrari1,ferrari2}.
In the $I_{0}^{*}$ case,  the states listed in Table $3$ all appear in the
stable spectrum at different points in moduli space.  It follows that the 
multiplets of the exceptional groups $E_{7}$ and $E_{6}$ 
into which these states are 
embedded also exist in the spectrum. However, nothing is known about the 
stability of the multiplets which decouple upon deforming the exceptional 
fibers into $I_0^*$.

\vskip 0.2in
{\bf Acknowledgements}

We would like to thank P. Candelas, D. Luest, D. Morrison and B. Zwiebach
for useful conversations and discussions.
B. Ovrut is supported in part by a Senior Alexander von Humboldt Award,
by the DOE under contract No. DE-ACO2-76-ER-03071 and by the University of
Pennsylvania Research Foundation Grant. Z. Guralnick is supported in part
by  the DOE under contract No. DE-ACO2-76-ER-03071.
A. Grassi is supported in part by an NSF grant DMS-9706707.

\section{Appendix A-- Review of String Junctions:}

This Appendix contains a review of string junctions, and
is a brief summary of work found in \cite{bartonone,bartontwo}.
There, string junctions were discussed in the context of type IIB 
string theory on ${\mathbb {CP}}^1$. As described in this paper, 
membrane junctions in an
elliptically fibered surface in M--theory become string junctions when 
projected into the base.

The equivalence classes of string junctions form a lattice
with a quadratic form.  To define the junction lattice, one initially 
splits all Kodaira fibers into a number of type $I_1$ fibers by a 
suitable deformation of the Weierstrass model (a ``relevant 
deformation"). The string junctions lie in the base space. 
The $I_1$ loci are points in the base and the string junctions start
from a fixed point $P$ away from these loci and may end 
only at the $I_1$ points.
Each segment of the junction is an oriented string with charges $(p,q)$,
which are conserved at branching points. 
The charges $(p,q)$ correspond to a one--cycle in the associated elliptic
fiber $F$ over the point $P$. The segment of a string 
junction ending at an $I_1$ locus  
has charges which are proportional to the vanishing cycle $(p_i,q_i)$ 
for the $i$-th $I_1$ fiber. The vanishing cycle is the eigenvector of the monodromy
matrix
\begin{equation}
\cM_i = 
\begin{pmatrix} 1-p_iq_i & p^2_i \\
		-q_i^2 & 1+p_iq_i \end{pmatrix}
\label{eq:A1}
\end{equation}
In a convenient decomposition \cite{bartonone},  any Kodaira fiber may be
split into $I_1$ fibers on the real axis each with vanishing cycles
$(1,0)$, $(1,-1)$, or $(1,1)$. The associated monodromy matrices are,
using~\eqref{eq:A1}, given by
\begin{equation}
A=  \begin{pmatrix}
                  1 & 1 \\
                  0 & 1
                  \end{pmatrix},
\qquad 
B=  \begin{pmatrix}
                  2 & 1 \\
                  -1 & 0
                  \end{pmatrix},
\qquad
C=  \begin{pmatrix}
                  0 & 1 \\
                  -1 & 2
                  \end{pmatrix}
\label{eq:burt}
\end{equation}
respectively.
The decompositions of the general Kodaira type fibers into $I_1$ fibers, 
from left to right on the real axis, are listed below.

\vspace{30pt}

\begin{tabular}{|l|c|} 
\hline
Kodaira type & decomposition\\
\hline
$I_n$ & $ A^n$ \\
\hline
$II$ & $AB$ \\
\hline
$III$ & $AAB$ \\
\hline
$IV$ & $AAAB$ \\
\hline
$I_n^*$ & $A^{4+n}BC$ \\
\hline
$IV^*$ & $AAAAABCC$\\
\hline
$III^*$ & $AAAAAABCC$ \\
\hline
$II^*$ & $AAAAAAABCC$ \\
\hline
\end{tabular}

\vspace{30pt}

Table 4:  The decomposition of Kodaira fibers into type $I_{1}$ fibers and the
associated monodromy.

\vspace{30pt}

The basis cycles on the elliptic fiber are not globally defined, and there
are branch cuts which may be taken to extend vertically downward from each 
$I_1$ loci.  As a segment of the string junction crosses the branch cut 
of the $i$-th $I_1$ locus, the $(p_{i},q_{i})$ charges labeling the cycle 
over this segment are acted on by the monodromy $\cM_i$. 
Such a segment can be pulled across the $I_1$ locus so that it
no longer crosses the branch cut.  However,  because of a Hanany-Witten
effect, an additional segment appears stretching between the original segment 
and the $I_1$ locus,  as illustrated in Figure $8$.
The charges of the new segment are determined by charge conservation to be 
$(p^{\prime},q^{\prime}) = \cM_i(p,q) - (p,q)$, which is proportional to
the vanishing cycle $(p_i, q_i)$. The two junctions related by pulling this 
segment across the $I_1$ locus 
are equivalent.  Thus, the equivalence classes of junctions can be 
determined by considering only junctions which have no components below
the real axis,  as in Figure $8(b)$.\footnote{Alternatively: 
One would like to write the vanishing cycle $(p,q)$ for any path $\Gamma$, 
which starts from $P$ and ends at one $I_1$ point $Q$, as a suitable
sum of our basic vanishing cycles. This sum should uniquely
identify a class of string junctions in the base. 
In fact, $\Gamma$  can be decomposed
into a product of simple loops around each $I_1$  point and a simple
path to $Q$ (that is a path which is homotopically trivial
in the complement of the $I_1$ points in the base). The composition
of the corresponding monodromy matrices applied to the cycle
$(p,q)$ is a multiple  of the vanishing cycle for the point $Q$.
Now because $\cM_i(r,s) - (r,s)$ is always proportional to
the vanishing cycle $(p_i, q_i)$, for all $i$ and $ (r,s)$, we can  show that the vanishing cycle for the path $\Gamma$ has
the expected form.}  The equivalence classes are then
labeled by vectors $\vec{J} = \sum_{i} n_i\vec{v}_i$,
where the integers $n_i$ indicate that the 
charges of the segment ending on the $i$-th $I_1$ locus are $n_i(p_i,q_i)$. 
The dimension of the string junction lattice is equal to the number of 
$I_1$ fibers.  The reader is referred to \cite{bartonone} for details 
on the construction of the quadratic intersection form on the junction
lattice. Here, we simply state the
results. The intersection form is given by
\begin{equation}
\vec{v}_i \cdot \vec{v}_i = -1
\label{eq:AA}
\end{equation}
\begin{equation}
\vec{v}_i \cdot \vec{v}_j = \frac{1}{2} (p_iq_j - q_ip_j), \quad i\ne j
\label{eq:A2}
\end{equation}

The basis in which an equivalence class $\vec{J}$ is labeled by the 
integers $n_i$ is not the most useful one.  
There is another basis in which the gauge and global symmetry charges 
appear explicitly. Viewed as quantum numbers of a BPS state, 
the electric and magnetic charges of 
a string junction are $(p,q) = \sum_i n_i(p_i,q_i)$.  The remaining directions in
the lattice are related to global symmetry charges. 
Using the intersection form on the lattice, one can show that the junction
lattice contains the weight lattice of a (simply laced) Lie algebra
\cite{bartonone}.  
The simple roots of the Lie algebra correspond to a basis set of string junctions
with vanishing charges $(p,q)=(0,0)$, and have self--intersection $-2$.
The intersection matrix of the simple root junctions $\vec{\alpha}_i$ 
is minus the Cartan matrix of the Lie algebra
\begin{equation}
\vec{\alpha}_i \cdot \vec{\alpha}_j = -C_{ij}.
\label{eq:A3}
\end{equation}
When $I_1$ fibers coalesce to form another Kodaira fiber, some root junctions 
have vanishing length. The Lie algebra associated with these simple roots 
generates a global symmetry of the theory which 
matches the A-D-E type of the Kodaira fiber. 
Fundamental weight junctions $\vec{w}^i$ with vanishing $(p,q)$ charges  are defined
by $\vec{w}^i \cdot \vec{\alpha}_j  = -\delta^i_j$, and are  
related to the simple root junctions by
\begin{equation}
\vec{w}^i = C^{-1ij}\vec{\alpha}_j.
\label{eq:A4}
\end{equation}
The fundamental weight junctions are ``improper'' in the sense that  
the $n_i$ associated with them are not integers.
To get a complete basis,  one defines another pair of string junctions
$\vec{w}_p$ and $\vec{w}_q$. These are orthogonal to the fundamental weight
junctions $\vec{w}^i$
and carry total charges $(p,q) = (1,0)$ and $(0,1)$ respectively.
The junctions $\vec{w}_p$ and $\vec{w}_q$ are also improper.
Any proper junction $\vec{J}$ can be written as
\begin{equation}
\vec{J} = p\vec{w}_p + q\vec{w}_q + a_i\vec{w}^i.
\label{eq:A5}
\end{equation}
where the integers $a_i$ are the Dynkin labels.
The weights in a BPS representation consist of junctions $\vec{J}$
related by the addition of simple roots, and satisfying the BPS 
condition \cite{bartontwo,sethi}
\begin{equation}
\vec {J} \cdot \vec{J} \ge -2 + gcd (p,q),
\label{eq:A6}
\end{equation}
where $gcd$ denotes the greatest common positive divisor of $p$ and $q$.

\onefigure{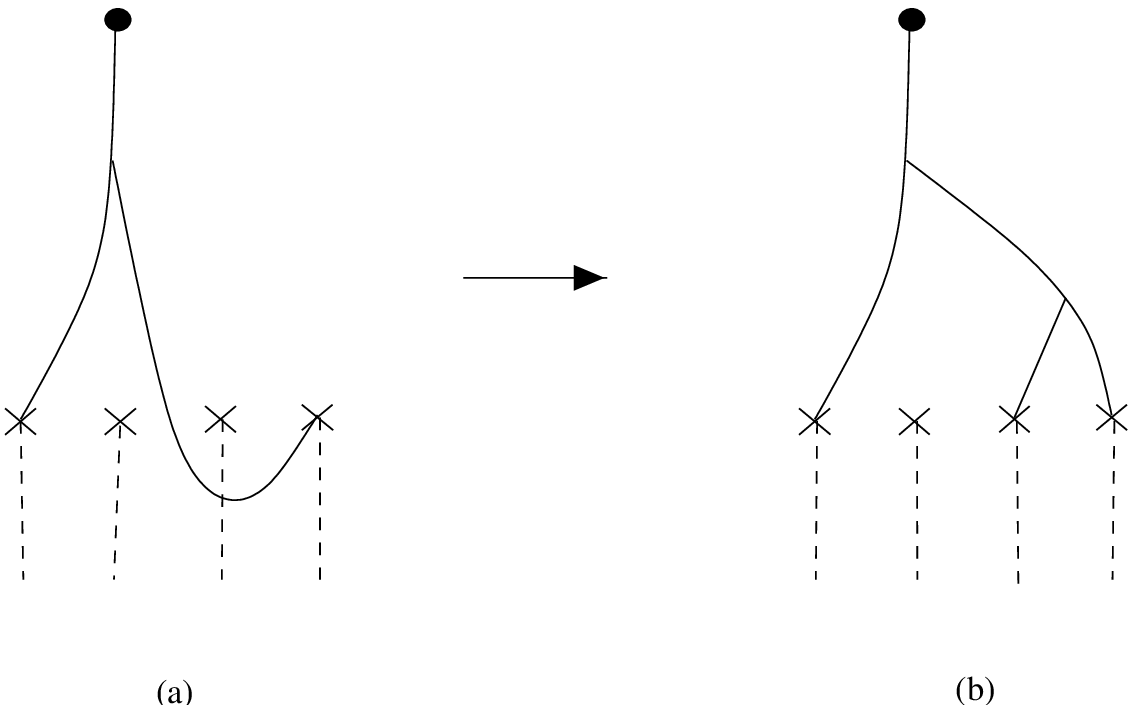}{Two equivalent string junction lattices. 
The dashed lines indicate branch cuts.}

\end{document}